\documentclass[aps,prl,twocolumn,superscriptaddress]{revtex4-1}
\usepackage{longtable}
\usepackage{amsfonts}
\usepackage{amsmath}
\usepackage{graphicx}
\usepackage{stackengine,graphicx}
\usepackage{epstopdf}
\usepackage{xcolor}
\usepackage[caption=false]{subfig}
\usepackage[hidelinks]{hyperref}
\definecolor{darkblue}{rgb}{0,0,0.55}
\hypersetup{
	colorlinks=true,
	citecolor=darkblue,
	linkcolor=darkblue
}

\begin{document}

 \title{Interaction-induced crossover between weak anti-localization and weak localization in a disordered InAs/GaSb double quantum well}

\author{Vahid Sazgari}
\affiliation{Faculty of Engineering and Natural Sciences, Sabanci University, Tuzla, 34956 Istanbul, Turkey}
\affiliation{Sabanci University Nanotechnology Research and Application Center, Tuzla, 34956 Istanbul, Turkey}
\author{Gerard Sullivan}
\affiliation{Teledyne Scientific and Imaging, Thousand Oaks, CA 91630, USA}
\author{\.{I}smet \.{I} Kaya}
\affiliation{Faculty of Engineering and Natural Sciences, Sabanci University, Tuzla, 34956 Istanbul, Turkey}
\affiliation{Sabanci University Nanotechnology Research and Application Center, Tuzla, 34956 Istanbul, Turkey}

 \email{iikaya@sabanciuniv.edu}

 \begin{abstract}

We present magneto-transport study in an InAs/GaSb double quantum well structure in the weak localization regime. As the charge carriers are depleted using a top gate electrode, we observe a crossover from weak anti-localization (WAL) to weak localization (WL), when the inelastic phase breaking time decreases below spin-orbit characteristic time as a result of enhanced electron-electron interactions at lower carrier concentrations. The same crossover is observed with increasing temperature. The linear temperature behavior of inelastic scattering rate indicates that the dominant phase breaking mechanism in our 2D system is due to electron-electron interaction. 
\end{abstract}

\date{\today}
\maketitle
InAs/GaSb heterostructures have recently been revisited by condensed matter physicists in quest for observing a topological phase of matter namely the quantum spin Hall insulator(QSHI)~\cite{Liu2008,KnezPRL107,KnezPRL109,SuzukiPRB2013,KnezPRL112,NichelePRL2014,KnezPRL2014,SpantonPRL2014,DuPRL2015,QuPRL2015,PribiagNatNano2015,LiPRL115,KaralicPRB2016,NicheleNJP18,NguyenPRL117,Yu_2018,ShojaeiPRM2,PhysRevB.97.245419,PhysRevB.99.085307,PhysRevB.99.201402,PRLMay2019,Sazgari}. Topological insulators inherently possess strong spin-orbit coupling (SOC) together with band inversion~\cite{TI,Spintronics2}. Strong
spin-orbit coupling embodied in InAs/GaSb bilayer QW
system has been comprehensively studied both experimentally and theoretically. SOC in crystal structures stems from inversion asymmetry in the crystal potential which builds up an internal electric field. Analogous to the electric field of nuclei, the crystal field generates an effective magnetic field in the rest frame of the moving electrons, which in turn couples the spin and momentum of the electron ~\cite{winkler2003spin,Spintronics-SOI}. This coupling can be in the form of Dresselhaus interaction caused by a bulk inversion asymmetry~\cite{dresselhaus1955spin}, e.g. in zinc blende structures, or Rashba spin-orbit contribution due to structural inversion asymmetry of the confinement potential~\cite{RashbaSOI2,RashbaSOI,Rashba4,Rashba1,Rashba2,Rashba3,RashbaSOI3}.
The spin-orbit interaction lifts the spin degeneracy at even a zero magnetic field and provides a versatile tool to manipulate the electron spin without an external magnetic field, which is desired for spintronic applications and quantum information technology~\cite{Spintronics1,Spintronics2,Spintronics3}. For example, classical spin Hall effect and its inverse driven by SOC are useful tools for injection and detection of the spin in spintronics~\cite{SHE}.

The effect of spin-orbit interaction has been widely studied in several low dimensional systems including two-dimensional electron gases (2DEGs)~\cite{RashbaStrength2,WL4,WL5,WL6,WL3}, QW heterostructures~\cite{WL8,RashbaStrength3,RashbaStrength4,RashbaStrength5,RashbaStrength6,RashbaAbsent,RashbaStrong,RashbaGate,Nitta,BeatSdH3,Beukman_BeatPattern,BeatSdH2,Nichele_BeatingPattern,WL2}, topological insulators~\cite{WL7,WL1}, etc. A strong SOC can be reflected in magnetoresistance measurements in two different ways. In high mobility systems a beating pattern in SdH oscillations may be observed as a result of two different SdH frequencies corresponding to carrier densities of each spin orientation split by SOC~\cite{RashbaGate,BeatSdH3,Beukman_BeatPattern,BeatSdH2,Nichele_BeatingPattern}. On the other hand, in disordered low mobility structures, the strength of spin-orbit interaction can be quantified by the spin relaxation length ($L_{so}$) obtained from the low field magnetoresistance measurements where the strong SOC appears as a weak anti-localization phenomenon~\cite{WL2,WL1}.   

In weakly disordered electronic systems, various scattering mechanisms in competition with each other determine the transport behavior of mesoscopic devices~\cite{HLN,WL-HLN,Yasuhiro}. Corresponding scattering rates or equivalently characteristic lengths, including the spin relaxation length $L_{so}$, can be extracted from low-field magnetotransport study of the longitudinal conductivity. In a spin-degenerate system, the quantum interference of the electrons moving on time-reversed closed paths is constructive and gives rise to a negative correction to conductivity This phenomenon is called weak localization (WL) and a small magnetic field can suppress the the excess resistance due to this quantum interference. Nevertheless, in presence of SOC, the quantum correction to the zero-field conductivity becomes positive which is manifested by a decreased conductivity upon application of a small magnetic field known as the weak anti-localization (WAL) effect.

In this work, we report a gate-tunable and temperature-induced crossover from WAL to WL. Inelastic scattering rate due to electron-electron interactions is enhanced at low electron densities and leads to reduction of the dephasing length, $L_{\phi}$. By applying a top gate voltage, $V_{tg}$ we were able to tune the electron density and consequently the electron dephasing length to go below the SOC length.  The enhancement of the inelastic scattering at lower densities implies that electron-electron interactions are the dominant inelastic scattering mechanism. However, the spin relaxation, $L_{so}$ and elastic scattering, $L_{e}$ lengths remain independent of $V_{tg}$. The WAL-WL crossover is observed also in the magnetoconductance plots as the temperature is varied; at elevated temperatures $L_{\phi}$ decreases below $L_{so}$ transitioning from WAL regime at low temperatures to WL regime at high temperatures.

The heterostructure used in the experiments was grown on a GaAs substrate by molecular beam epitaxy. Following an AlSb buffer layer of 1~$\mu m$ thickness, the bilayer QW structure consisting of a 12.5~nm InAs layer on a 10~nm GaSb layer confined between two 50~nm thick AlGaSb barriers was grown. The top barrier was protected from oxidization by a 3~nm GaSb cap layer. A 20$\times$5~$\mu$m ($L \times W$) Hall bar device was used in the measurements. The device mesa was formed by standard electron beam lithography followed by chemical wet etching. The etched surfaces were passivated by a 200~nm thick PECVD-grown Si$_3$Ni$_4$ layer which also served as the dielectric layer separating the top gate electrodes (Ti/Au) from the heterostructure. The ohmic contacts were made by Ge/Au/Ni metalization without annealing. Transport measurements were performed in a dilution refrigerator with base temperature of 10~mK using standard lock-in techniques with 10~nA current excitation at 7~Hz unless otherwise stated.

\begin{figure}
	\centering \includegraphics[width=0.95\columnwidth]{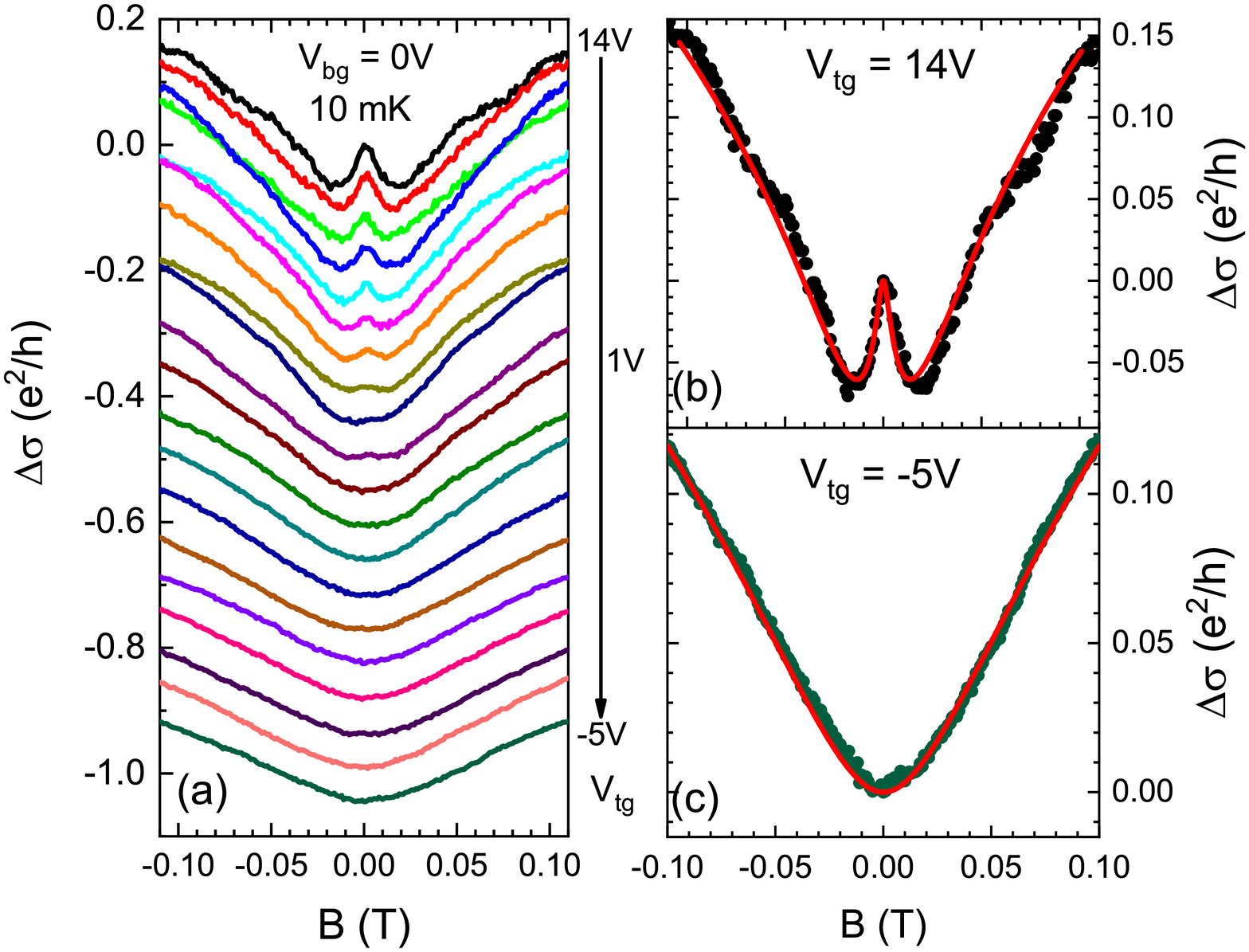}
	\includegraphics[width=0.95\columnwidth]{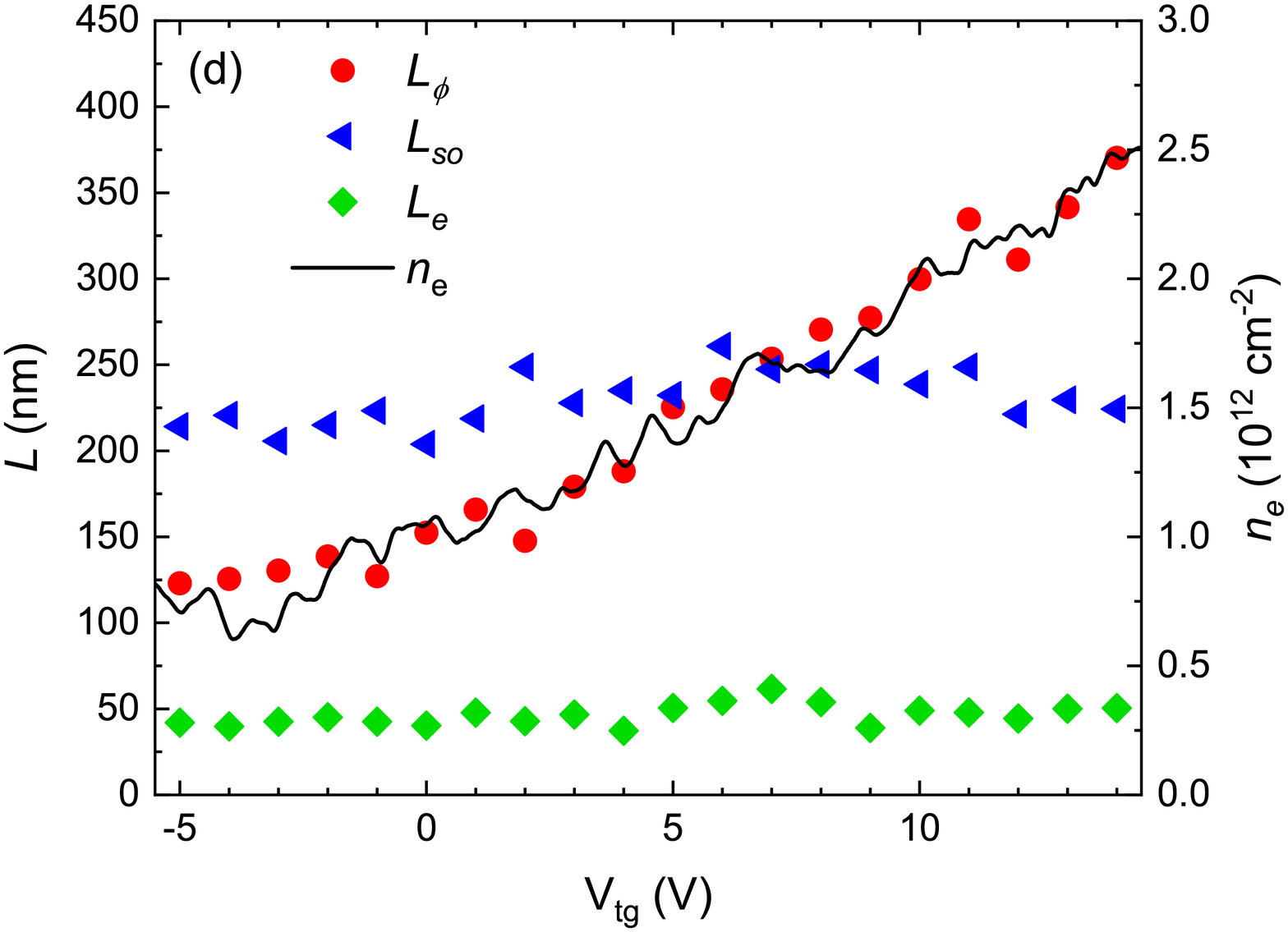} 
	\caption{Magnetotransport measurements and the calculated characteristic scattering lengths. (a) Change in the conductance with magnetic field B at different $V_{tg}$'s where the WAL behavior at $V_{tg} = 14$~V is gradually suppressed as the gate voltage is lowered and transitions into WL around $V_{tg} = 4$~V.  (b) and (c) Change in conductance versus B at $V_{tg}=+14$~V and $V_{tg}=-5$~V, respectively, that exhibit WAL and WL features with the fit curves (red) obtained from the HLN equation. (d) Characteristic lengths involved in WL and WAL as functions of $V_{tg}$ obtained from the HLN fits to the experimental data. The density modulation $n_e$ vs $V_{tg}$, which is calculated from the Hall  measurements is also shown.}\label{FIG:W1864-WAL-gate}
\end{figure} 

\begin{figure}[t]
		\centering
		\includegraphics[width=0.45\textwidth]{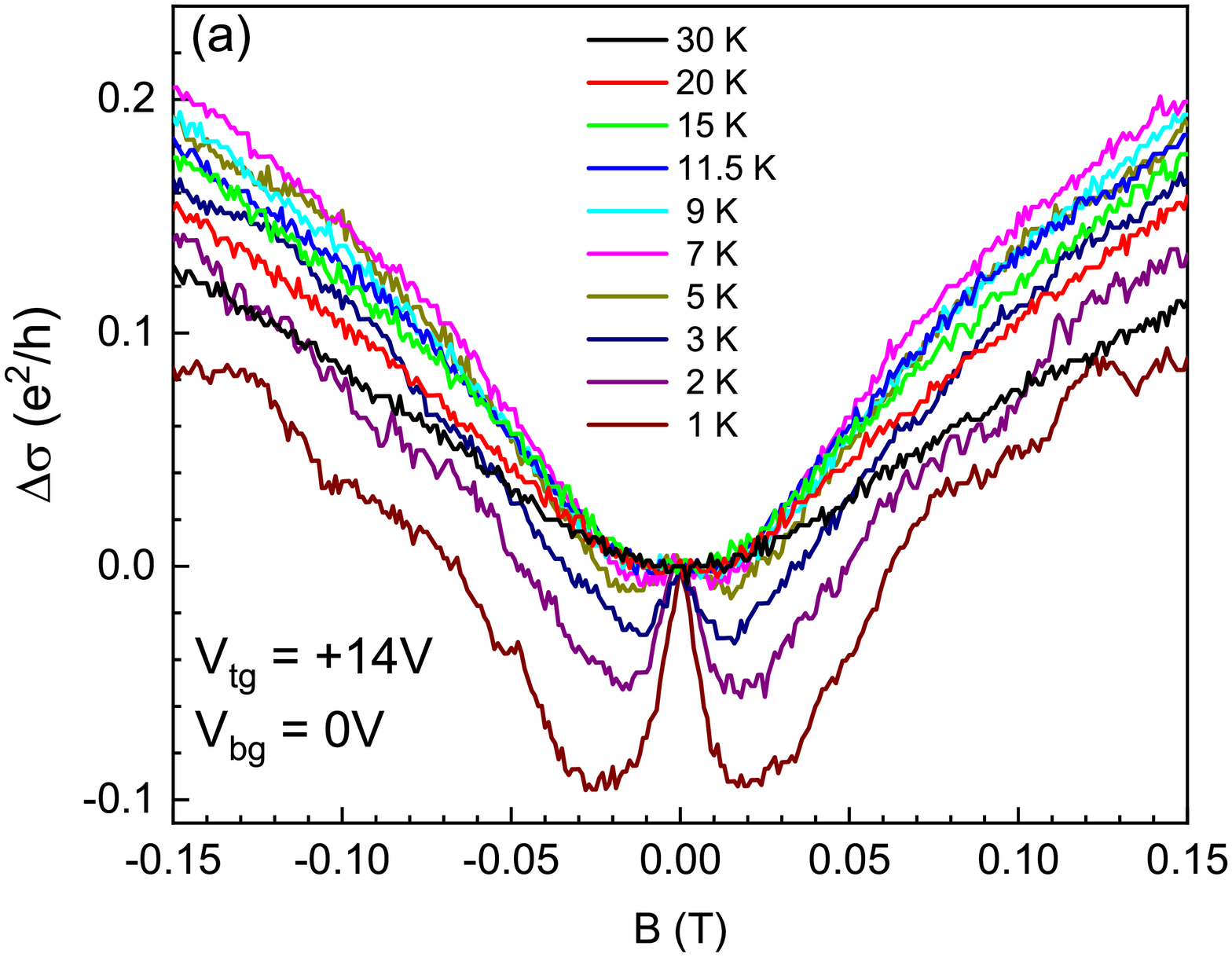}
		\includegraphics[width=0.45\textwidth]{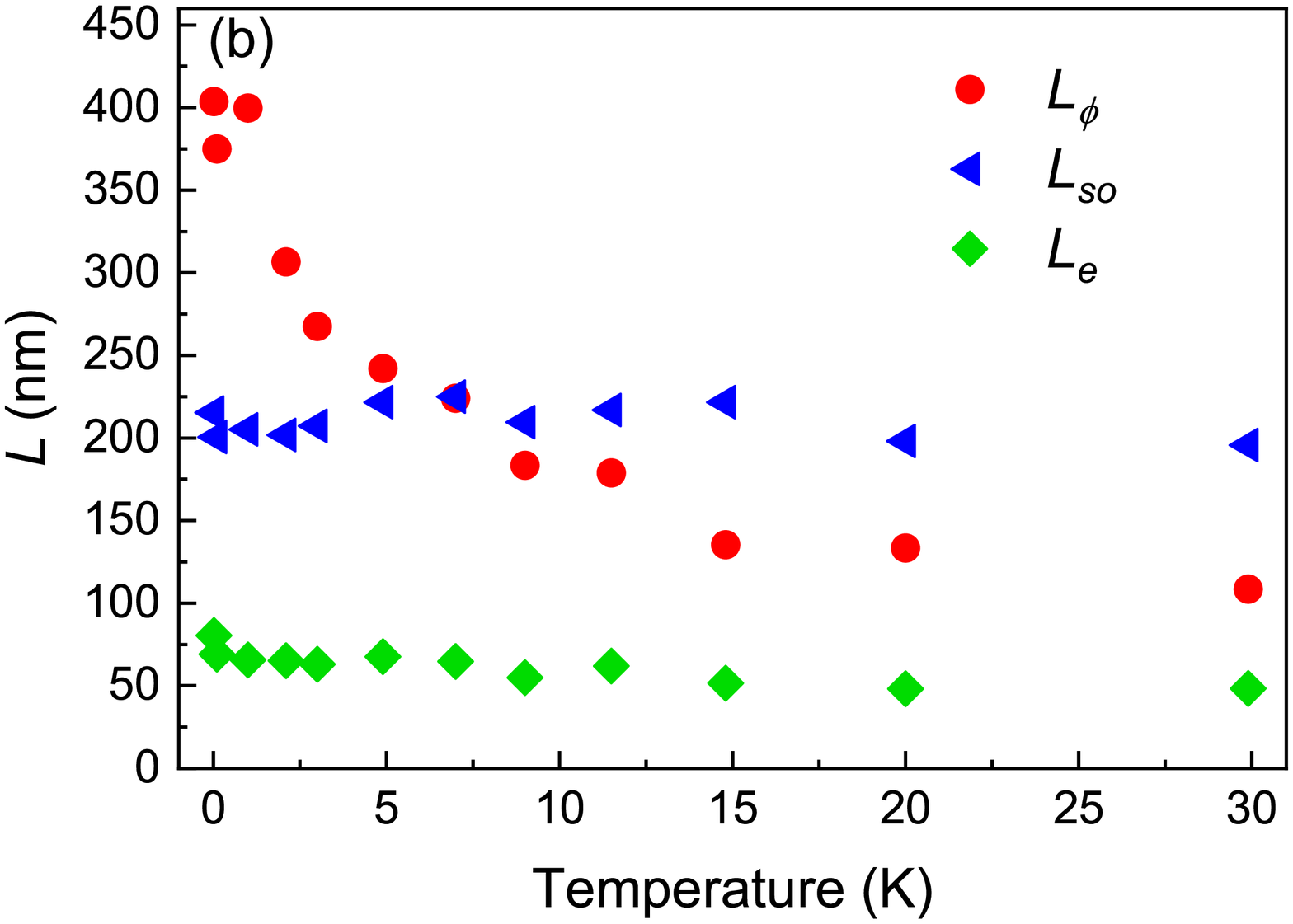}
		\caption{Temperature-induced WAL/WL crossover. (a) Magnetoconductance quantum correction measured at different temperatures when $V_{bg}=0V$ and $V_{tg}=+14V$. (b) The inelastic dephasing length $L_{\phi}$, spin relaxation length $L_{so}$, and elastic mean free path $L_{e}$ as functions of temperature obtained from theoretical fits to the measurements shown in (a).}\label{FIG:W1864-WAL-Temp}

\end{figure}    

Similar to the heterostructure we used in our recent report on the localization of trivial edge states~\cite{Sazgari}, the structure in this study is also disordered by single layer of silicon atoms, however, in different position along the growth direction. In Ref.~\cite{Sazgari}, Si atoms were deposited inside the InAs at 2 atomic layer distance from the InAs/GaSb interface, while in the heterostructure of this work, the silicon atoms were placed at 1 atomic layer distance into the InAs layer  with an average density of $10^{11}$ cm$^{-2}$. Difference in the delta doping location resulted in a remarkable change in the transport properties which is possibly due to impurity-induced interfacial effects. In our previous study, the bilayer InAs/GaSb heterostructure exhibited a trivial insulator behavior with a sizable gap using dual gate electrodes. However, in the present work, the bilayer QW shows a semi-metallic behavior within the whole range of top and bottom gate voltages~\cite{Suppl}. 

The hallmark of the present study is the observation of a crossover between WAL and WL regimes in the magneto-resistance measurements that can be induced either by the top gate voltage or temperature. The magnetotransport measurements are well described by Hikami– Larkin– Nagaoka (HLN) interference model~\cite{HLN} which characterizes the WL and WAL phenomena in disordered two dimensional electronic systems by different scattering mechanisms and their corresponding length scales. According to HLN model, the quantum interference correction to the low field magnetoconductivity is given by~\cite{HLN,WL-HLN}:

\begin{align} 
\Delta \sigma (B)& = -\frac{e^2}{2\pi h} \Big\{\psi\left(\frac{B_{\phi}}{B}+\frac{1}{2}\right) +2\psi\left(\frac{B_{so}+B_{e}}{B}+\frac{1}{2}\right) \nonumber\\-&\quad 3\psi\left(\frac{\left(4/3\right)B_{so}+B_{\phi}}{B}+\frac{1}{2}\right) -\ln\left(\frac{B_{\phi}}{B}\right)\nonumber\\-&\quad2\ln\left(\frac{B_{so}+B_{e}}{B}\right)+3\ln\left(\frac{\left(4/3\right)B_{so}+B_{\phi}}{B}\right)\Big\},
\end{align}\label{EQN:HLN}


\noindent where $\psi(x)$ is the digamma function and $B_{i}$'s are the characteristic magnetic fields each corresponds to a different scattering mechanism. By fitting the above equation to the magnetoconductance data we obtained the characteristic fields for inelastic dephasing $B_{\phi}$, spin-orbit $B_{so}$ and elastic $B_{e}$ scatterings. One can then obtain the corresponding scattering length via the relation $B_{i}=\hbar/\left(4eL_{i}^2\right)$. 

By application of the top gate voltage, a continuous change from WAL to WL regime is observed in the magnetotransport measurements. Figure~\ref{FIG:W1864-WAL-gate} illustrates the change of the magneto-conductivity for different top gate voltages from +14~V to -5~V with 1~V steps. While at higher gate voltages the device represents a WAL feature, it gradually transforms into WL regime as the inelastic dephasing length $L_{\phi}$ declines by lowering the top gate voltage and finally crosses over the spin-orbit scattering length $L_{so}$ at $V_{tg}\sim5V$ where the WAL cusp diminishes and the sample acquires a WL feature  (Fig.~\ref{FIG:W1864-WAL-gate}(a). The most pronounced WAL and WL characteristics at extreme gate voltages are illustrated in Fig.~\ref{FIG:W1864-WAL-gate}b and \ref{FIG:W1864-WAL-gate}c, respectively, with perfect fits to the HLN equation. The scattering lengths calculated from  each magnetoconductance curve are plotted as functions of top gate voltage in Fig.~\ref{FIG:W1864-WAL-gate}d. The elastic scattering length and the spin-orbit length are nearly independent of gate voltage, however, it can be seen that the inelastic dephasing length is strongly gate-dependent and decreases by decreasing the gate voltage. To demonstrate the correlation between $L_{\phi}$ and the carrier density $n_{e}$, we plotted the electron density versus top gate voltage in the same graph. It is visible that the dephasing length due to inelastic scattering is linked to the carrier density suggesting that the dominant inelastic process is induced by the electron-electron interactions.
It can also be seen in Fig.~\ref{FIG:W1864-WAL-gate}(d) that the density modulation in response to the top gate electric field is slightly nonlinear. This is attributed to surface trap states common to the III-V semiconductor interfaces~\cite{Sazgari,GateHysteresis,GateHysteresis2} or impurity-induced charge states inside the bilayer QW. 

This may explain the gate-independent spin relaxation length $L_{so}$ which is otherwise expected to change with the external electric field via the Rashba spin-orbit interaction~\cite{WL8,RashbaStrength2,RashbaStrength3,RashbaStrength4,RashbaStrength5,RashbaStrength6,RashbaAbsent,RashbaStrong,RashbaGate,BeatSdH3,Beukman_BeatPattern,WL2}. In other words, the out of plane electric field in the conducting channel is dominated by the crystal fields rather than the external gate-induced field. However, for 2D electron systems, the reported strengths of the Rashba effect are markedly different~\cite{WL8,RashbaStrength2,RashbaStrength3,RashbaStrength4,RashbaStrength5,RashbaStrength6,RashbaAbsent,RashbaStrong,RashbaGate}. There is a puzzling discrepancy between theoretical expectations and experimental observations of the effect of external electric field on the Rashba SOI such that it is sometimes very weak or even absent~\cite{RashbaAbsent} and in some experiments far stronger than what theory predicts~\cite{RashbaStrong,RashbaGate}. For instance in Ref.~\cite{RashbaStrong}, a weak perpendicular electric field (2 orders of magnitude smaller than the typical built-in electric field) applied to the plane of a 2D hole gas was able to change the Rashba SOI by $20\%$, whereas in reference to \cite{RashbaGate}, the Rashba coupling was enhanced by two folds when the field was changed only by $10\%$ within the range of applied gate voltage. Therefore, the exact effect of gate-induced electric field on the Rashba SOI has remained ambiguous. Above all, a precise investigation of Rashba effect incorporates the variation of electric field at constant carrier densities because otherwise the effects of density modulation and electric field variation on the amount of Rashba SOI can not be separated.   

On the other hand, we observe the electron dephasing length to be a function of carrier concentration. At low temperatures in disordered electronic systems, the phase breaking length is determined by electron-electron collisions in the form of direct Coulomb interaction between conducting electrons or inelastic scattering by the electromagnetic field fluctuations in the potential landscape generated by the moving electrons. The e-e inelastic scattering is sensitive to the carrier concentration as it changes the potential landscape for cruising electrons. At low temperatures, the scattering rate corresponding to the e-e collisions with small energy transfer, often referred to as Nyquist scattering, is given by~\cite{Altshuler,Taboryski_1990}:

\begin{equation} \label{EQN:NiquistTime}
\tau_{N} = \frac{2E_{F}\tau_{t}}{k_{B}T\ln\left(E_{F}\tau_{t}/\hbar\right)} \equiv \frac{\hbar g}{k_{B}T\ln\left(g/2\right)},
\end{equation}

\noindent where, $E_{F}=\hbar^2 k_{F}^2/2m^{\star}$ is the Fermi energy, $m^{\star}$ is the electron effective mass, and $\tau_{t}=\mu m^{\star}/e$ is the transport time also called mean free time i.e. the time between consecutive scattering events. And $g$ is the dimensionless conductivity in units of $e^2/h$. The characteristic scattering lengths are related to their corresponding times via $L_i^2 = D \tau_i$, where $D$ is the electron diffusion coefficient given by:

\begin{equation} \label{EQN:Diffusion}
D = \frac{\hbar^2 k_F^2 \tau_t}{2 m^2} \equiv \frac{\hbar g}{2m^\star},
\end{equation}

Referring to Eq.~\ref{EQN:NiquistTime}, at a given temperature, the Nyquist time $\tau_{N}$ thus the dephasing length decreases at lower charge densities $n_{e}$ (or equivalently smaller $E_{F}$) consistent with our observation of gate-dependent dephasing length.   
As illustrated in Fig.~\ref{FIG:W1864-WAL-gate}(d), we observe that $L_{\phi}$ reduces monotonically with decreasing electron concentration $n_{e}$ at lower top gate voltages in agreement with the Nyquist interaction model. Therefore, in the heterostructure of this study, the phase breaking time $\tau_{\phi}$ is determined by the characteristic time of the electron scattering by electromagnetic fluctuations with small energy transfer that is the Nyquist time ($\tau_{\phi}\sim \tau_{N}$).

\begin{figure}
	\centering \includegraphics[width=0.95\columnwidth]{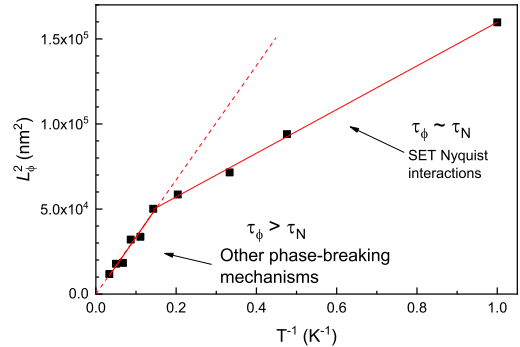} 
	\caption{$L_{\phi}^{2}\propto \tau_{\phi}$ shows a linear behavior with inverse temperature ($T^{-1}$) as expected for inelastic phase breaking scatterings induced by e-e interactions. At low temperatures the SET Nyquist scattering is dominated while at higher temperatures the effective dephasing time is larger than the Nyquist time.}\label{FIG:L2-inverseT}
\end{figure}

Theory predicts that inelastic scattering rates magnify at higher temperatures. The temperature dependence of the quantum interference corrections due to electron scatterings is quantified by the time $\tau_{\phi}$ during which the electron wavefunction remains phase-coherent. In highly interacting two-dimensional electron systems, the phase breaking scattering mechanisms are dominated by Nyquist scattering rate induced by electron-electron interaction given by Eq.~\ref{EQN:NiquistTime}. We conducted the temperature study of the magnetoconductance at $V_{tg}=+14V$ where a prominent WAL behavior is observed at lowest temperatures which indicates that the elastic and spin-orbit scatterings are dominant over inelastic phase breaking interactions. As expected, the inelastic scattering mechanisms strengthen at increased temperatures leading to a transition from WAL to WL regime when the $L_{\phi}$ reduces below the $L_{so}$. The temperature study of the magnetoconductance correction is indicated in Fig.~\ref{FIG:W1864-WAL-Temp}. The positive cusp of the conductance correction around zero field, that characterizes the WAL, gradually suppresses by increasing the temperature until it enters the WL regime at $T\approx7$~K where the crossover occurs (Fig.~\ref{FIG:W1864-WAL-Temp}a). The extracted characteristic lengths which quantifies the crossover are shown in Fig.~\ref{FIG:W1864-WAL-Temp}b where $L_{e}$ and $L_{so}$ are insensitive to temperature. Consistent with Eq.~\ref{EQN:NiquistTime}, $L_{\phi}$, which is here equivalent to $L_{N}$, decreases with rising temperature. As illustrated in Fig.~\ref{FIG:L2-inverseT}, the square of the dephasing length ($L_{\phi}^2\propto \tau_{N}$) is linearly proportional to the inverse temperature. However, the slopes in the WL and WAL regimes are different. At sufficiently low temperatures ($T~\le 7$~K), where the WAL effect is dominant, the slope of the fit line is $1.28~(\pm 0.03) \times 10^{-13}$~m$^2$K. On the other hand, substituting all the numbers (assuming the effective mass as $m = 0.04~m_{e}$~\cite{Beukman_BeatPattern}) in equation ~\ref{EQN:NiquistTime}, which is valid for low temperatures, we obtain a slope of $\approx1.25 \times 10^{-13}$~m$^2$K which is in perfect agreement with the line slope for WAL (low temperature) regime in Fig.~\ref{FIG:L2-inverseT}. At higher temperatures, however, in the WL regime, the slope at which the dephasing length (the Nyquist scattering rate) decreases (increases) with temperature is higher (lower) than that for low temperature regime. The slope of the fit line for $T \ge 7$K is $3.4~(\pm 0.4) \times 10^{-13}$~m$^2$K more than two fold larger than that for lower temperatures. At elevated temperatures, the inelastic dephasing mechanisms may deviate from small energy transfer Nyquist interactions due to thermal excitations when the larger energy transfer collisions may play the dominant role. Indeed the phase breaking electron-electron interactions reduces at higher temperatures and the inelastic dephasing rate $\tau_{\phi}^{-1}$ is decreased by the logarithmic factor in Eq.~\ref{EQN:NiquistTime}~\cite{Altshuler}.      

In conclusion, we observed an interaction-induced crossover between the WAL and WL phenomena in our intentionally disordered InAs/GaSb bilayer QW structure driven by top gate voltage and temperature. The heterostructure with Si delta-doped atoms near the interface of electron and hole QWs demonstrates WAL feature in magnetoconductance measurements at high carrier densities at lowest temperatures as an evidence for strong SOI. Using the theoretical HLN model, we obtained the gate and temperature dependences of the characteristic scattering lengths corresponding to three different scattering mechanisms. The elastic scattering length is found as  $L_{e}\sim 50$~nm which is consistent with the mean spacing between Si dopants of $\sim 30$~nm. It is almost independent of gate voltage and temperature, which indicates the short range scattering induced by impurity or interface roughness is dominant. The SO scattering length also shows no dependence on temperature and gate voltage. We believe that the effective gate-induced electric field is reduced by impurity and trap states leading to a gate-independent $L_{so}$. The only parameter responsible for the temperature and gate-driven crossover is the electron dephasing length $L_{\phi}$ which changes the magnetotransport behavior when it crosses over $L_{so}$ with an indication of enhanced $e-e$ interactions at lower temperatures and charge concentrations. The temperature and gate dependent analysis of the inelastic phase breaking length indicates that the Nyquist scattering, characterized by electron-electron interactions with SET, is the dominant dephasing mechanism at low temperatures where the spin-orbit scattering is still governing the magneto-conductance behavior with a WAL feature. At higher temperatures the inealastic scattering takes over the SO interaction thus leading to the WL effect. At the same time, the dephasing time increases above the Nyquist time meaning that the Nyquist interactions suppressed at higher temperatures and other inelastic processes perhaps with larger energy transfer becomes important. 

\bibliographystyle{apsrev4-1}
\bibliography{Main-WAL_WL}

\begin{thebibliography}{66}%
\makeatletter
\providecommand \@ifxundefined [1]{%
 \@ifx{#1\undefined}
}%
\providecommand \@ifnum [1]{%
 \ifnum #1\expandafter \@firstoftwo
 \else \expandafter \@secondoftwo
 \fi
}%
\providecommand \@ifx [1]{%
 \ifx #1\expandafter \@firstoftwo
 \else \expandafter \@secondoftwo
 \fi
}%
\providecommand \natexlab [1]{#1}%
\providecommand \enquote  [1]{``#1''}%
\providecommand \bibnamefont  [1]{#1}%
\providecommand \bibfnamefont [1]{#1}%
\providecommand \citenamefont [1]{#1}%
\providecommand \href@noop [0]{\@secondoftwo}%
\providecommand \href [0]{\begingroup \@sanitize@url \@href}%
\providecommand \@href[1]{\@@startlink{#1}\@@href}%
\providecommand \@@href[1]{\endgroup#1\@@endlink}%
\providecommand \@sanitize@url [0]{\catcode `\\12\catcode `\$12\catcode
  `\&12\catcode `\#12\catcode `\^12\catcode `\_12\catcode `\%12\relax}%
\providecommand \@@startlink[1]{}%
\providecommand \@@endlink[0]{}%
\providecommand \url  [0]{\begingroup\@sanitize@url \@url }%
\providecommand \@url [1]{\endgroup\@href {#1}{\urlprefix }}%
\providecommand \urlprefix  [0]{URL }%
\providecommand \Eprint [0]{\href }%
\providecommand \doibase [0]{http://dx.doi.org/}%
\providecommand \selectlanguage [0]{\@gobble}%
\providecommand \bibinfo  [0]{\@secondoftwo}%
\providecommand \bibfield  [0]{\@secondoftwo}%
\providecommand \translation [1]{[#1]}%
\providecommand \BibitemOpen [0]{}%
\providecommand \bibitemStop [0]{}%
\providecommand \bibitemNoStop [0]{.\EOS\space}%
\providecommand \EOS [0]{\spacefactor3000\relax}%
\providecommand \BibitemShut  [1]{\csname bibitem#1\endcsname}%
\let\auto@bib@innerbib\@empty
\bibitem [{\citenamefont {Liu}\ \emph {et~al.}(2008)\citenamefont {Liu},
  \citenamefont {Hughes}, \citenamefont {Qi}, \citenamefont {Wang},\ and\
  \citenamefont {Zhang}}]{Liu2008}%
  \BibitemOpen
  \bibfield  {author} {\bibinfo {author} {\bibfnamefont {C.}~\bibnamefont
  {Liu}}, \bibinfo {author} {\bibfnamefont {T.~L.}\ \bibnamefont {Hughes}},
  \bibinfo {author} {\bibfnamefont {X.-L.}\ \bibnamefont {Qi}}, \bibinfo
  {author} {\bibfnamefont {K.}~\bibnamefont {Wang}}, \ and\ \bibinfo {author}
  {\bibfnamefont {S.-C.}\ \bibnamefont {Zhang}},\ }\href {\doibase
  10.1103/PhysRevLett.100.236601} {\bibfield  {journal} {\bibinfo  {journal}
  {Phys. Rev. Lett.}\ }\textbf {\bibinfo {volume} {100}},\ \bibinfo {pages}
  {236601} (\bibinfo {year} {2008})}\BibitemShut {NoStop}%
\bibitem [{\citenamefont {Knez}\ \emph {et~al.}(2011)\citenamefont {Knez},
  \citenamefont {Du},\ and\ \citenamefont {Sullivan}}]{KnezPRL107}%
  \BibitemOpen
  \bibfield  {author} {\bibinfo {author} {\bibfnamefont {I.}~\bibnamefont
  {Knez}}, \bibinfo {author} {\bibfnamefont {R.-R.}\ \bibnamefont {Du}}, \ and\
  \bibinfo {author} {\bibfnamefont {G.}~\bibnamefont {Sullivan}},\ }\href
  {\doibase 10.1103/PhysRevLett.107.136603} {\bibfield  {journal} {\bibinfo
  {journal} {Phys. Rev. Lett.}\ }\textbf {\bibinfo {volume} {107}},\ \bibinfo
  {pages} {136603} (\bibinfo {year} {2011})}\BibitemShut {NoStop}%
\bibitem [{\citenamefont {Knez}\ \emph {et~al.}(2012)\citenamefont {Knez},
  \citenamefont {Du},\ and\ \citenamefont {Sullivan}}]{KnezPRL109}%
  \BibitemOpen
  \bibfield  {author} {\bibinfo {author} {\bibfnamefont {I.}~\bibnamefont
  {Knez}}, \bibinfo {author} {\bibfnamefont {R.-R.}\ \bibnamefont {Du}}, \ and\
  \bibinfo {author} {\bibfnamefont {G.}~\bibnamefont {Sullivan}},\ }\href
  {\doibase 10.1103/PhysRevLett.109.186603} {\bibfield  {journal} {\bibinfo
  {journal} {Phys. Rev. Lett.}\ }\textbf {\bibinfo {volume} {109}},\ \bibinfo
  {pages} {186603} (\bibinfo {year} {2012})}\BibitemShut {NoStop}%
\bibitem [{\citenamefont {Suzuki}\ \emph {et~al.}(2013)\citenamefont {Suzuki},
  \citenamefont {Harada}, \citenamefont {Onomitsu},\ and\ \citenamefont
  {Muraki}}]{SuzukiPRB2013}%
  \BibitemOpen
  \bibfield  {author} {\bibinfo {author} {\bibfnamefont {K.}~\bibnamefont
  {Suzuki}}, \bibinfo {author} {\bibfnamefont {Y.}~\bibnamefont {Harada}},
  \bibinfo {author} {\bibfnamefont {K.}~\bibnamefont {Onomitsu}}, \ and\
  \bibinfo {author} {\bibfnamefont {K.}~\bibnamefont {Muraki}},\ }\href
  {\doibase 10.1103/PhysRevB.87.235311} {\bibfield  {journal} {\bibinfo
  {journal} {Phys. Rev. B}\ }\textbf {\bibinfo {volume} {87}},\ \bibinfo
  {pages} {235311} (\bibinfo {year} {2013})}\BibitemShut {NoStop}%
\bibitem [{\citenamefont {Knez}\ \emph
  {et~al.}(2014{\natexlab{a}})\citenamefont {Knez}, \citenamefont {Rettner},
  \citenamefont {Yang}, \citenamefont {Parkin}, \citenamefont {Du},
  \citenamefont {Du},\ and\ \citenamefont {Sullivan}}]{KnezPRL112}%
  \BibitemOpen
  \bibfield  {author} {\bibinfo {author} {\bibfnamefont {I.}~\bibnamefont
  {Knez}}, \bibinfo {author} {\bibfnamefont {C.~T.}\ \bibnamefont {Rettner}},
  \bibinfo {author} {\bibfnamefont {S.-H.}\ \bibnamefont {Yang}}, \bibinfo
  {author} {\bibfnamefont {S.~S.~P.}\ \bibnamefont {Parkin}}, \bibinfo {author}
  {\bibfnamefont {L.}~\bibnamefont {Du}}, \bibinfo {author} {\bibfnamefont
  {R.-R.}\ \bibnamefont {Du}}, \ and\ \bibinfo {author} {\bibfnamefont
  {G.}~\bibnamefont {Sullivan}},\ }\href {\doibase
  10.1103/PhysRevLett.112.026602} {\bibfield  {journal} {\bibinfo  {journal}
  {Phys. Rev. Lett.}\ }\textbf {\bibinfo {volume} {112}},\ \bibinfo {pages}
  {026602} (\bibinfo {year} {2014}{\natexlab{a}})}\BibitemShut {NoStop}%
\bibitem [{\citenamefont {Nichele}\ \emph {et~al.}(2014)\citenamefont
  {Nichele}, \citenamefont {Pal}, \citenamefont {Pietsch}, \citenamefont {Ihn},
  \citenamefont {Ensslin}, \citenamefont {Charpentier},\ and\ \citenamefont
  {Wegscheider}}]{NichelePRL2014}%
  \BibitemOpen
  \bibfield  {author} {\bibinfo {author} {\bibfnamefont {F.}~\bibnamefont
  {Nichele}}, \bibinfo {author} {\bibfnamefont {A.~N.}\ \bibnamefont {Pal}},
  \bibinfo {author} {\bibfnamefont {P.}~\bibnamefont {Pietsch}}, \bibinfo
  {author} {\bibfnamefont {T.}~\bibnamefont {Ihn}}, \bibinfo {author}
  {\bibfnamefont {K.}~\bibnamefont {Ensslin}}, \bibinfo {author} {\bibfnamefont
  {C.}~\bibnamefont {Charpentier}}, \ and\ \bibinfo {author} {\bibfnamefont
  {W.}~\bibnamefont {Wegscheider}},\ }\href {\doibase
  10.1103/PhysRevLett.112.036802} {\bibfield  {journal} {\bibinfo  {journal}
  {Phys. Rev. Lett.}\ }\textbf {\bibinfo {volume} {112}},\ \bibinfo {pages}
  {036802} (\bibinfo {year} {2014})}\BibitemShut {NoStop}%
\bibitem [{\citenamefont {Knez}\ \emph
  {et~al.}(2014{\natexlab{b}})\citenamefont {Knez}, \citenamefont {Rettner},
  \citenamefont {Yang}, \citenamefont {Parkin}, \citenamefont {Du},
  \citenamefont {Du},\ and\ \citenamefont {Sullivan}}]{KnezPRL2014}%
  \BibitemOpen
  \bibfield  {author} {\bibinfo {author} {\bibfnamefont {I.}~\bibnamefont
  {Knez}}, \bibinfo {author} {\bibfnamefont {C.~T.}\ \bibnamefont {Rettner}},
  \bibinfo {author} {\bibfnamefont {S.-H.}\ \bibnamefont {Yang}}, \bibinfo
  {author} {\bibfnamefont {S.~S.~P.}\ \bibnamefont {Parkin}}, \bibinfo {author}
  {\bibfnamefont {L.}~\bibnamefont {Du}}, \bibinfo {author} {\bibfnamefont
  {R.-R.}\ \bibnamefont {Du}}, \ and\ \bibinfo {author} {\bibfnamefont
  {G.}~\bibnamefont {Sullivan}},\ }\href {\doibase
  10.1103/PhysRevLett.112.026602} {\bibfield  {journal} {\bibinfo  {journal}
  {Phys. Rev. Lett.}\ }\textbf {\bibinfo {volume} {112}},\ \bibinfo {pages}
  {026602} (\bibinfo {year} {2014}{\natexlab{b}})}\BibitemShut {NoStop}%
\bibitem [{\citenamefont {Spanton}\ \emph {et~al.}(2014)\citenamefont
  {Spanton}, \citenamefont {Nowack}, \citenamefont {Du}, \citenamefont
  {Sullivan}, \citenamefont {Du},\ and\ \citenamefont
  {Moler}}]{SpantonPRL2014}%
  \BibitemOpen
  \bibfield  {author} {\bibinfo {author} {\bibfnamefont {E.~M.}\ \bibnamefont
  {Spanton}}, \bibinfo {author} {\bibfnamefont {K.~C.}\ \bibnamefont {Nowack}},
  \bibinfo {author} {\bibfnamefont {L.}~\bibnamefont {Du}}, \bibinfo {author}
  {\bibfnamefont {G.}~\bibnamefont {Sullivan}}, \bibinfo {author}
  {\bibfnamefont {R.-R.}\ \bibnamefont {Du}}, \ and\ \bibinfo {author}
  {\bibfnamefont {K.~A.}\ \bibnamefont {Moler}},\ }\href {\doibase
  10.1103/PhysRevLett.113.026804} {\bibfield  {journal} {\bibinfo  {journal}
  {Phys. Rev. Lett.}\ }\textbf {\bibinfo {volume} {113}},\ \bibinfo {pages}
  {026804} (\bibinfo {year} {2014})}\BibitemShut {NoStop}%
\bibitem [{\citenamefont {Du}\ \emph {et~al.}(2015)\citenamefont {Du},
  \citenamefont {Knez}, \citenamefont {Sullivan},\ and\ \citenamefont
  {Du}}]{DuPRL2015}%
  \BibitemOpen
  \bibfield  {author} {\bibinfo {author} {\bibfnamefont {L.}~\bibnamefont
  {Du}}, \bibinfo {author} {\bibfnamefont {I.}~\bibnamefont {Knez}}, \bibinfo
  {author} {\bibfnamefont {G.}~\bibnamefont {Sullivan}}, \ and\ \bibinfo
  {author} {\bibfnamefont {R.-R.}\ \bibnamefont {Du}},\ }\href {\doibase
  10.1103/PhysRevLett.114.096802} {\bibfield  {journal} {\bibinfo  {journal}
  {Phys. Rev. Lett.}\ }\textbf {\bibinfo {volume} {114}},\ \bibinfo {pages}
  {096802} (\bibinfo {year} {2015})}\BibitemShut {NoStop}%
\bibitem [{\citenamefont {Qu}\ \emph {et~al.}(2015)\citenamefont {Qu},
  \citenamefont {Beukman}, \citenamefont {Nadj-Perge}, \citenamefont {Wimmer},
  \citenamefont {Nguyen}, \citenamefont {Yi}, \citenamefont {Thorp},
  \citenamefont {Sokolich}, \citenamefont {Kiselev}, \citenamefont {Manfra},
  \citenamefont {Marcus},\ and\ \citenamefont {Kouwenhoven}}]{QuPRL2015}%
  \BibitemOpen
  \bibfield  {author} {\bibinfo {author} {\bibfnamefont {F.}~\bibnamefont
  {Qu}}, \bibinfo {author} {\bibfnamefont {A.~J.~A.}\ \bibnamefont {Beukman}},
  \bibinfo {author} {\bibfnamefont {S.}~\bibnamefont {Nadj-Perge}}, \bibinfo
  {author} {\bibfnamefont {M.}~\bibnamefont {Wimmer}}, \bibinfo {author}
  {\bibfnamefont {B.-M.}\ \bibnamefont {Nguyen}}, \bibinfo {author}
  {\bibfnamefont {W.}~\bibnamefont {Yi}}, \bibinfo {author} {\bibfnamefont
  {J.}~\bibnamefont {Thorp}}, \bibinfo {author} {\bibfnamefont
  {M.}~\bibnamefont {Sokolich}}, \bibinfo {author} {\bibfnamefont {A.~A.}\
  \bibnamefont {Kiselev}}, \bibinfo {author} {\bibfnamefont {M.~J.}\
  \bibnamefont {Manfra}}, \bibinfo {author} {\bibfnamefont {C.~M.}\
  \bibnamefont {Marcus}}, \ and\ \bibinfo {author} {\bibfnamefont {L.~P.}\
  \bibnamefont {Kouwenhoven}},\ }\href {\doibase
  10.1103/PhysRevLett.115.036803} {\bibfield  {journal} {\bibinfo  {journal}
  {Phys. Rev. Lett.}\ }\textbf {\bibinfo {volume} {115}},\ \bibinfo {pages}
  {036803} (\bibinfo {year} {2015})}\BibitemShut {NoStop}%
\bibitem [{\citenamefont {Pribiag}\ \emph {et~al.}(2015)\citenamefont
  {Pribiag}, \citenamefont {Beukman}, \citenamefont {Qu}, \citenamefont
  {Cassidy}, \citenamefont {Charpentier}, \citenamefont {Wegscheider},\ and\
  \citenamefont {Kouwenhoven}}]{PribiagNatNano2015}%
  \BibitemOpen
  \bibfield  {author} {\bibinfo {author} {\bibfnamefont {V.~S.}\ \bibnamefont
  {Pribiag}}, \bibinfo {author} {\bibfnamefont {A.~J.~A.}\ \bibnamefont
  {Beukman}}, \bibinfo {author} {\bibfnamefont {F.}~\bibnamefont {Qu}},
  \bibinfo {author} {\bibfnamefont {M.~C.}\ \bibnamefont {Cassidy}}, \bibinfo
  {author} {\bibfnamefont {C.}~\bibnamefont {Charpentier}}, \bibinfo {author}
  {\bibfnamefont {W.}~\bibnamefont {Wegscheider}}, \ and\ \bibinfo {author}
  {\bibfnamefont {L.~P.}\ \bibnamefont {Kouwenhoven}},\ }\href {\doibase
  10.1038/nnano.2015.86} {\bibfield  {journal} {\bibinfo  {journal} {Nature
  Nanotechnology}\ }\textbf {\bibinfo {volume} {10}},\ \bibinfo {pages} {593}
  (\bibinfo {year} {2015})}\BibitemShut {NoStop}%
\bibitem [{\citenamefont {Li}\ \emph {et~al.}(2015)\citenamefont {Li},
  \citenamefont {Wang}, \citenamefont {Fu}, \citenamefont {Du}, \citenamefont
  {Schreiber}, \citenamefont {Mu}, \citenamefont {Liu}, \citenamefont
  {Sullivan}, \citenamefont {Cs\'athy}, \citenamefont {Lin},\ and\
  \citenamefont {Du}}]{LiPRL115}%
  \BibitemOpen
  \bibfield  {author} {\bibinfo {author} {\bibfnamefont {T.}~\bibnamefont
  {Li}}, \bibinfo {author} {\bibfnamefont {P.}~\bibnamefont {Wang}}, \bibinfo
  {author} {\bibfnamefont {H.}~\bibnamefont {Fu}}, \bibinfo {author}
  {\bibfnamefont {L.}~\bibnamefont {Du}}, \bibinfo {author} {\bibfnamefont
  {K.~A.}\ \bibnamefont {Schreiber}}, \bibinfo {author} {\bibfnamefont
  {X.}~\bibnamefont {Mu}}, \bibinfo {author} {\bibfnamefont {X.}~\bibnamefont
  {Liu}}, \bibinfo {author} {\bibfnamefont {G.}~\bibnamefont {Sullivan}},
  \bibinfo {author} {\bibfnamefont {G.~A.}\ \bibnamefont {Cs\'athy}}, \bibinfo
  {author} {\bibfnamefont {X.}~\bibnamefont {Lin}}, \ and\ \bibinfo {author}
  {\bibfnamefont {R.-R.}\ \bibnamefont {Du}},\ }\href {\doibase
  10.1103/PhysRevLett.115.136804} {\bibfield  {journal} {\bibinfo  {journal}
  {Phys. Rev. Lett.}\ }\textbf {\bibinfo {volume} {115}},\ \bibinfo {pages}
  {136804} (\bibinfo {year} {2015})}\BibitemShut {NoStop}%
\bibitem [{\citenamefont {Karalic}\ \emph {et~al.}(2016)\citenamefont
  {Karalic}, \citenamefont {Mueller}, \citenamefont {Mittag}, \citenamefont
  {Pakrouski}, \citenamefont {Wu}, \citenamefont {Soluyanov}, \citenamefont
  {Troyer}, \citenamefont {Tschirky}, \citenamefont {Wegscheider},
  \citenamefont {Ensslin},\ and\ \citenamefont {Ihn}}]{KaralicPRB2016}%
  \BibitemOpen
  \bibfield  {author} {\bibinfo {author} {\bibfnamefont {M.}~\bibnamefont
  {Karalic}}, \bibinfo {author} {\bibfnamefont {S.}~\bibnamefont {Mueller}},
  \bibinfo {author} {\bibfnamefont {C.}~\bibnamefont {Mittag}}, \bibinfo
  {author} {\bibfnamefont {K.}~\bibnamefont {Pakrouski}}, \bibinfo {author}
  {\bibfnamefont {Q.}~\bibnamefont {Wu}}, \bibinfo {author} {\bibfnamefont
  {A.~A.}\ \bibnamefont {Soluyanov}}, \bibinfo {author} {\bibfnamefont
  {M.}~\bibnamefont {Troyer}}, \bibinfo {author} {\bibfnamefont
  {T.}~\bibnamefont {Tschirky}}, \bibinfo {author} {\bibfnamefont
  {W.}~\bibnamefont {Wegscheider}}, \bibinfo {author} {\bibfnamefont
  {K.}~\bibnamefont {Ensslin}}, \ and\ \bibinfo {author} {\bibfnamefont
  {T.}~\bibnamefont {Ihn}},\ }\href {\doibase 10.1103/PhysRevB.94.241402}
  {\bibfield  {journal} {\bibinfo  {journal} {Phys. Rev. B}\ }\textbf {\bibinfo
  {volume} {94}},\ \bibinfo {pages} {241402} (\bibinfo {year}
  {2016})}\BibitemShut {NoStop}%
\bibitem [{\citenamefont {Nichele}\ \emph {et~al.}(2016)\citenamefont
  {Nichele}, \citenamefont {Suominen}, \citenamefont {Kjaergaard},
  \citenamefont {Marcus}, \citenamefont {Sajadi}, \citenamefont {Folk},
  \citenamefont {Qu}, \citenamefont {Beukman}, \citenamefont {de~Vries},
  \citenamefont {van Veen}, \citenamefont {Nadj-Perge}, \citenamefont
  {Kouwenhoven}, \citenamefont {Nguyen}, \citenamefont {Kiselev}, \citenamefont
  {Yi}, \citenamefont {Sokolich}, \citenamefont {Manfra}, \citenamefont
  {Spanton},\ and\ \citenamefont {Moler}}]{NicheleNJP18}%
  \BibitemOpen
  \bibfield  {author} {\bibinfo {author} {\bibfnamefont {F.}~\bibnamefont
  {Nichele}}, \bibinfo {author} {\bibfnamefont {H.~J.}\ \bibnamefont
  {Suominen}}, \bibinfo {author} {\bibfnamefont {M.}~\bibnamefont
  {Kjaergaard}}, \bibinfo {author} {\bibfnamefont {C.~M.}\ \bibnamefont
  {Marcus}}, \bibinfo {author} {\bibfnamefont {E.}~\bibnamefont {Sajadi}},
  \bibinfo {author} {\bibfnamefont {J.~A.}\ \bibnamefont {Folk}}, \bibinfo
  {author} {\bibfnamefont {F.}~\bibnamefont {Qu}}, \bibinfo {author}
  {\bibfnamefont {A.~J.~A.}\ \bibnamefont {Beukman}}, \bibinfo {author}
  {\bibfnamefont {F.~K.}\ \bibnamefont {de~Vries}}, \bibinfo {author}
  {\bibfnamefont {J.}~\bibnamefont {van Veen}}, \bibinfo {author}
  {\bibfnamefont {S.}~\bibnamefont {Nadj-Perge}}, \bibinfo {author}
  {\bibfnamefont {L.~P.}\ \bibnamefont {Kouwenhoven}}, \bibinfo {author}
  {\bibfnamefont {B.-M.}\ \bibnamefont {Nguyen}}, \bibinfo {author}
  {\bibfnamefont {A.~A.}\ \bibnamefont {Kiselev}}, \bibinfo {author}
  {\bibfnamefont {W.}~\bibnamefont {Yi}}, \bibinfo {author} {\bibfnamefont
  {M.}~\bibnamefont {Sokolich}}, \bibinfo {author} {\bibfnamefont {M.~J.}\
  \bibnamefont {Manfra}}, \bibinfo {author} {\bibfnamefont {E.~M.}\
  \bibnamefont {Spanton}}, \ and\ \bibinfo {author} {\bibfnamefont {K.~A.}\
  \bibnamefont {Moler}},\ }\href
  {http://stacks.iop.org/1367-2630/18/i=8/a=083005} {\bibfield  {journal}
  {\bibinfo  {journal} {New Journal of Physics}\ }\textbf {\bibinfo {volume}
  {18}},\ \bibinfo {pages} {083005} (\bibinfo {year} {2016})}\BibitemShut
  {NoStop}%
\bibitem [{\citenamefont {Nguyen}\ \emph {et~al.}(2016)\citenamefont {Nguyen},
  \citenamefont {Kiselev}, \citenamefont {Noah}, \citenamefont {Yi},
  \citenamefont {Qu}, \citenamefont {Beukman}, \citenamefont {de~Vries},
  \citenamefont {van Veen}, \citenamefont {Nadj-Perge}, \citenamefont
  {Kouwenhoven}, \citenamefont {Kjaergaard}, \citenamefont {Suominen},
  \citenamefont {Nichele}, \citenamefont {Marcus}, \citenamefont {Manfra},\
  and\ \citenamefont {Sokolich}}]{NguyenPRL117}%
  \BibitemOpen
  \bibfield  {author} {\bibinfo {author} {\bibfnamefont {B.-M.}\ \bibnamefont
  {Nguyen}}, \bibinfo {author} {\bibfnamefont {A.~A.}\ \bibnamefont {Kiselev}},
  \bibinfo {author} {\bibfnamefont {R.}~\bibnamefont {Noah}}, \bibinfo {author}
  {\bibfnamefont {W.}~\bibnamefont {Yi}}, \bibinfo {author} {\bibfnamefont
  {F.}~\bibnamefont {Qu}}, \bibinfo {author} {\bibfnamefont {A.~J.~A.}\
  \bibnamefont {Beukman}}, \bibinfo {author} {\bibfnamefont {F.~K.}\
  \bibnamefont {de~Vries}}, \bibinfo {author} {\bibfnamefont {J.}~\bibnamefont
  {van Veen}}, \bibinfo {author} {\bibfnamefont {S.}~\bibnamefont
  {Nadj-Perge}}, \bibinfo {author} {\bibfnamefont {L.~P.}\ \bibnamefont
  {Kouwenhoven}}, \bibinfo {author} {\bibfnamefont {M.}~\bibnamefont
  {Kjaergaard}}, \bibinfo {author} {\bibfnamefont {H.~J.}\ \bibnamefont
  {Suominen}}, \bibinfo {author} {\bibfnamefont {F.}~\bibnamefont {Nichele}},
  \bibinfo {author} {\bibfnamefont {C.~M.}\ \bibnamefont {Marcus}}, \bibinfo
  {author} {\bibfnamefont {M.~J.}\ \bibnamefont {Manfra}}, \ and\ \bibinfo
  {author} {\bibfnamefont {M.}~\bibnamefont {Sokolich}},\ }\href {\doibase
  10.1103/PhysRevLett.117.077701} {\bibfield  {journal} {\bibinfo  {journal}
  {Phys. Rev. Lett.}\ }\textbf {\bibinfo {volume} {117}},\ \bibinfo {pages}
  {077701} (\bibinfo {year} {2016})}\BibitemShut {NoStop}%
\bibitem [{\citenamefont {Yu}\ \emph {et~al.}(2018)\citenamefont {Yu},
  \citenamefont {Cleric{\`{o}}}, \citenamefont {Fuentevilla}, \citenamefont
  {Shi}, \citenamefont {Jiang}, \citenamefont {Saha}, \citenamefont {Lou},
  \citenamefont {Chang}, \citenamefont {Huang}, \citenamefont {Gumbs},
  \citenamefont {Smirnov}, \citenamefont {Stanton}, \citenamefont {Jiang},
  \citenamefont {Bellani}, \citenamefont {Meziani}, \citenamefont {Diez},
  \citenamefont {Pan}, \citenamefont {Hawkins},\ and\ \citenamefont
  {Klem}}]{Yu_2018}%
  \BibitemOpen
  \bibfield  {author} {\bibinfo {author} {\bibfnamefont {W.}~\bibnamefont
  {Yu}}, \bibinfo {author} {\bibfnamefont {V.}~\bibnamefont {Cleric{\`{o}}}},
  \bibinfo {author} {\bibfnamefont {C.~H.}\ \bibnamefont {Fuentevilla}},
  \bibinfo {author} {\bibfnamefont {X.}~\bibnamefont {Shi}}, \bibinfo {author}
  {\bibfnamefont {Y.}~\bibnamefont {Jiang}}, \bibinfo {author} {\bibfnamefont
  {D.}~\bibnamefont {Saha}}, \bibinfo {author} {\bibfnamefont {W.~K.}\
  \bibnamefont {Lou}}, \bibinfo {author} {\bibfnamefont {K.}~\bibnamefont
  {Chang}}, \bibinfo {author} {\bibfnamefont {D.~H.}\ \bibnamefont {Huang}},
  \bibinfo {author} {\bibfnamefont {G.}~\bibnamefont {Gumbs}}, \bibinfo
  {author} {\bibfnamefont {D.}~\bibnamefont {Smirnov}}, \bibinfo {author}
  {\bibfnamefont {C.~J.}\ \bibnamefont {Stanton}}, \bibinfo {author}
  {\bibfnamefont {Z.}~\bibnamefont {Jiang}}, \bibinfo {author} {\bibfnamefont
  {V.}~\bibnamefont {Bellani}}, \bibinfo {author} {\bibfnamefont
  {Y.}~\bibnamefont {Meziani}}, \bibinfo {author} {\bibfnamefont
  {E.}~\bibnamefont {Diez}}, \bibinfo {author} {\bibfnamefont {W.}~\bibnamefont
  {Pan}}, \bibinfo {author} {\bibfnamefont {S.~D.}\ \bibnamefont {Hawkins}}, \
  and\ \bibinfo {author} {\bibfnamefont {J.~F.}\ \bibnamefont {Klem}},\ }\href
  {\doibase 10.1088/1367-2630/aac595} {\bibfield  {journal} {\bibinfo
  {journal} {New Journal of Physics}\ }\textbf {\bibinfo {volume} {20}},\
  \bibinfo {pages} {053062} (\bibinfo {year} {2018})}\BibitemShut {NoStop}%
\bibitem [{\citenamefont {Shojaei}\ \emph {et~al.}(2018)\citenamefont
  {Shojaei}, \citenamefont {McFadden}, \citenamefont {Pendharkar},
  \citenamefont {Lee}, \citenamefont {Flatt\'e},\ and\ \citenamefont
  {Palmstr\o{}m}}]{ShojaeiPRM2}%
  \BibitemOpen
  \bibfield  {author} {\bibinfo {author} {\bibfnamefont {B.}~\bibnamefont
  {Shojaei}}, \bibinfo {author} {\bibfnamefont {A.~P.}\ \bibnamefont
  {McFadden}}, \bibinfo {author} {\bibfnamefont {M.}~\bibnamefont
  {Pendharkar}}, \bibinfo {author} {\bibfnamefont {J.~S.}\ \bibnamefont {Lee}},
  \bibinfo {author} {\bibfnamefont {M.~E.}\ \bibnamefont {Flatt\'e}}, \ and\
  \bibinfo {author} {\bibfnamefont {C.~J.}\ \bibnamefont {Palmstr\o{}m}},\
  }\href {\doibase 10.1103/PhysRevMaterials.2.064603} {\bibfield  {journal}
  {\bibinfo  {journal} {Phys. Rev. Materials}\ }\textbf {\bibinfo {volume}
  {2}},\ \bibinfo {pages} {064603} (\bibinfo {year} {2018})}\BibitemShut
  {NoStop}%
\bibitem [{\citenamefont {Krishtopenko}\ \emph {et~al.}(2018)\citenamefont
  {Krishtopenko}, \citenamefont {Ruffenach}, \citenamefont {Gonzalez-Posada},
  \citenamefont {Boissier}, \citenamefont {Marcinkiewicz}, \citenamefont
  {Fadeev}, \citenamefont {Kadykov}, \citenamefont {Rumyantsev}, \citenamefont
  {Morozov}, \citenamefont {Gavrilenko}, \citenamefont {Consejo}, \citenamefont
  {Desrat}, \citenamefont {Jouault}, \citenamefont {Knap}, \citenamefont
  {Tourni\'e},\ and\ \citenamefont {Teppe}}]{PhysRevB.97.245419}%
  \BibitemOpen
  \bibfield  {author} {\bibinfo {author} {\bibfnamefont {S.~S.}\ \bibnamefont
  {Krishtopenko}}, \bibinfo {author} {\bibfnamefont {S.}~\bibnamefont
  {Ruffenach}}, \bibinfo {author} {\bibfnamefont {F.}~\bibnamefont
  {Gonzalez-Posada}}, \bibinfo {author} {\bibfnamefont {G.}~\bibnamefont
  {Boissier}}, \bibinfo {author} {\bibfnamefont {M.}~\bibnamefont
  {Marcinkiewicz}}, \bibinfo {author} {\bibfnamefont {M.~A.}\ \bibnamefont
  {Fadeev}}, \bibinfo {author} {\bibfnamefont {A.~M.}\ \bibnamefont {Kadykov}},
  \bibinfo {author} {\bibfnamefont {V.~V.}\ \bibnamefont {Rumyantsev}},
  \bibinfo {author} {\bibfnamefont {S.~V.}\ \bibnamefont {Morozov}}, \bibinfo
  {author} {\bibfnamefont {V.~I.}\ \bibnamefont {Gavrilenko}}, \bibinfo
  {author} {\bibfnamefont {C.}~\bibnamefont {Consejo}}, \bibinfo {author}
  {\bibfnamefont {W.}~\bibnamefont {Desrat}}, \bibinfo {author} {\bibfnamefont
  {B.}~\bibnamefont {Jouault}}, \bibinfo {author} {\bibfnamefont
  {W.}~\bibnamefont {Knap}}, \bibinfo {author} {\bibfnamefont {E.}~\bibnamefont
  {Tourni\'e}}, \ and\ \bibinfo {author} {\bibfnamefont {F.}~\bibnamefont
  {Teppe}},\ }\href {\doibase 10.1103/PhysRevB.97.245419} {\bibfield  {journal}
  {\bibinfo  {journal} {Phys. Rev. B}\ }\textbf {\bibinfo {volume} {97}},\
  \bibinfo {pages} {245419} (\bibinfo {year} {2018})}\BibitemShut {NoStop}%
\bibitem [{\citenamefont {Wu}\ \emph {et~al.}(2019)\citenamefont {Wu},
  \citenamefont {Lou}, \citenamefont {Chang}, \citenamefont {Sullivan},\ and\
  \citenamefont {Du}}]{PhysRevB.99.085307}%
  \BibitemOpen
  \bibfield  {author} {\bibinfo {author} {\bibfnamefont {X.}~\bibnamefont
  {Wu}}, \bibinfo {author} {\bibfnamefont {W.}~\bibnamefont {Lou}}, \bibinfo
  {author} {\bibfnamefont {K.}~\bibnamefont {Chang}}, \bibinfo {author}
  {\bibfnamefont {G.}~\bibnamefont {Sullivan}}, \ and\ \bibinfo {author}
  {\bibfnamefont {R.-R.}\ \bibnamefont {Du}},\ }\href {\doibase
  10.1103/PhysRevB.99.085307} {\bibfield  {journal} {\bibinfo  {journal} {Phys.
  Rev. B}\ }\textbf {\bibinfo {volume} {99}},\ \bibinfo {pages} {085307}
  (\bibinfo {year} {2019})}\BibitemShut {NoStop}%
\bibitem [{\citenamefont {Karalic}\ \emph {et~al.}(2019)\citenamefont
  {Karalic}, \citenamefont {Mittag}, \citenamefont {Mueller}, \citenamefont
  {Tschirky}, \citenamefont {Wegscheider}, \citenamefont {Ensslin},
  \citenamefont {Ihn},\ and\ \citenamefont {Glazman}}]{PhysRevB.99.201402}%
  \BibitemOpen
  \bibfield  {author} {\bibinfo {author} {\bibfnamefont {M.}~\bibnamefont
  {Karalic}}, \bibinfo {author} {\bibfnamefont {C.}~\bibnamefont {Mittag}},
  \bibinfo {author} {\bibfnamefont {S.}~\bibnamefont {Mueller}}, \bibinfo
  {author} {\bibfnamefont {T.}~\bibnamefont {Tschirky}}, \bibinfo {author}
  {\bibfnamefont {W.}~\bibnamefont {Wegscheider}}, \bibinfo {author}
  {\bibfnamefont {K.}~\bibnamefont {Ensslin}}, \bibinfo {author} {\bibfnamefont
  {T.}~\bibnamefont {Ihn}}, \ and\ \bibinfo {author} {\bibfnamefont
  {L.}~\bibnamefont {Glazman}},\ }\href {\doibase 10.1103/PhysRevB.99.201402}
  {\bibfield  {journal} {\bibinfo  {journal} {Phys. Rev. B}\ }\textbf {\bibinfo
  {volume} {99}},\ \bibinfo {pages} {201402} (\bibinfo {year}
  {2019})}\BibitemShut {NoStop}%
\bibitem [{\citenamefont {Xiao}\ \emph {et~al.}(2019)\citenamefont {Xiao},
  \citenamefont {Liu}, \citenamefont {Samarth},\ and\ \citenamefont
  {Hu}}]{PRLMay2019}%
  \BibitemOpen
  \bibfield  {author} {\bibinfo {author} {\bibfnamefont {D.}~\bibnamefont
  {Xiao}}, \bibinfo {author} {\bibfnamefont {C.-X.}\ \bibnamefont {Liu}},
  \bibinfo {author} {\bibfnamefont {N.}~\bibnamefont {Samarth}}, \ and\
  \bibinfo {author} {\bibfnamefont {L.-H.}\ \bibnamefont {Hu}},\ }\href
  {\doibase 10.1103/PhysRevLett.122.186802} {\bibfield  {journal} {\bibinfo
  {journal} {Phys. Rev. Lett.}\ }\textbf {\bibinfo {volume} {122}},\ \bibinfo
  {pages} {186802} (\bibinfo {year} {2019})}\BibitemShut {NoStop}%
\bibitem [{\citenamefont {Sazgari}\ \emph {et~al.}(2019)\citenamefont
  {Sazgari}, \citenamefont {Sullivan},\ and\ \citenamefont {Kaya}}]{Sazgari}%
  \BibitemOpen
  \bibfield  {author} {\bibinfo {author} {\bibfnamefont {V.}~\bibnamefont
  {Sazgari}}, \bibinfo {author} {\bibfnamefont {G.}~\bibnamefont {Sullivan}}, \
  and\ \bibinfo {author} {\bibfnamefont {I.~I.}\ \bibnamefont {Kaya}},\ }\href
  {\doibase 10.1103/PhysRevB.100.041404} {\bibfield  {journal} {\bibinfo
  {journal} {Phys. Rev. B}\ }\textbf {\bibinfo {volume} {100}},\ \bibinfo
  {pages} {041404} (\bibinfo {year} {2019})}\BibitemShut {NoStop}%
\bibitem [{\citenamefont {Hasan}\ and\ \citenamefont {Kane}(2010)}]{TI}%
  \BibitemOpen
  \bibfield  {author} {\bibinfo {author} {\bibfnamefont {M.~Z.}\ \bibnamefont
  {Hasan}}\ and\ \bibinfo {author} {\bibfnamefont {C.~L.}\ \bibnamefont
  {Kane}},\ }\href {\doibase 10.1103/RevModPhys.82.3045} {\bibfield  {journal}
  {\bibinfo  {journal} {Rev. Mod. Phys.}\ }\textbf {\bibinfo {volume} {82}},\
  \bibinfo {pages} {3045} (\bibinfo {year} {2010})}\BibitemShut {NoStop}%
\bibitem [{\citenamefont {Qi}\ and\ \citenamefont
  {Zhang}(2011)}]{Spintronics2}%
  \BibitemOpen
  \bibfield  {author} {\bibinfo {author} {\bibfnamefont {X.-L.}\ \bibnamefont
  {Qi}}\ and\ \bibinfo {author} {\bibfnamefont {S.-C.}\ \bibnamefont {Zhang}},\
  }\href {\doibase 10.1103/RevModPhys.83.1057} {\bibfield  {journal} {\bibinfo
  {journal} {Rev. Mod. Phys.}\ }\textbf {\bibinfo {volume} {83}},\ \bibinfo
  {pages} {1057} (\bibinfo {year} {2011})}\BibitemShut {NoStop}%
\bibitem [{\citenamefont {Winkler}(2003)}]{winkler2003spin}%
  \BibitemOpen
  \bibfield  {author} {\bibinfo {author} {\bibfnamefont {R.}~\bibnamefont
  {Winkler}},\ }\href@noop {} {\bibfield  {journal} {\bibinfo  {journal}
  {Springer Tracts in Modern Physics}\ }\textbf {\bibinfo {volume} {191}},\
  \bibinfo {pages} {1} (\bibinfo {year} {2003})}\BibitemShut {NoStop}%
\bibitem [{\citenamefont {Fabian}\ \emph {et~al.}(2007)\citenamefont {Fabian},
  \citenamefont {Matos-Abiague}, \citenamefont {Ertler}, \citenamefont
  {Stano},\ and\ \citenamefont {{\v{Z}}uti{\'c}}}]{Spintronics-SOI}%
  \BibitemOpen
  \bibfield  {author} {\bibinfo {author} {\bibfnamefont {J.}~\bibnamefont
  {Fabian}}, \bibinfo {author} {\bibfnamefont {A.}~\bibnamefont
  {Matos-Abiague}}, \bibinfo {author} {\bibfnamefont {C.}~\bibnamefont
  {Ertler}}, \bibinfo {author} {\bibfnamefont {P.}~\bibnamefont {Stano}}, \
  and\ \bibinfo {author} {\bibfnamefont {I.}~\bibnamefont {{\v{Z}}uti{\'c}}},\
  }\href@noop {} {\bibfield  {journal} {\bibinfo  {journal} {Acta Physica
  Slovaca. Reviews and Tutorials}\ }\textbf {\bibinfo {volume} {57}},\ \bibinfo
  {pages} {565} (\bibinfo {year} {2007})}\BibitemShut {NoStop}%
\bibitem [{\citenamefont {Dresselhaus}(1955)}]{dresselhaus1955spin}%
  \BibitemOpen
  \bibfield  {author} {\bibinfo {author} {\bibfnamefont {G.}~\bibnamefont
  {Dresselhaus}},\ }\href@noop {} {\bibfield  {journal} {\bibinfo  {journal}
  {Physical Review}\ }\textbf {\bibinfo {volume} {100}},\ \bibinfo {pages}
  {580} (\bibinfo {year} {1955})}\BibitemShut {NoStop}%
\bibitem [{\citenamefont {J.~Ohkawa}\ and\ \citenamefont
  {Uemura}(1974)}]{RashbaSOI2}%
  \BibitemOpen
  \bibfield  {author} {\bibinfo {author} {\bibfnamefont {F.}~\bibnamefont
  {J.~Ohkawa}}\ and\ \bibinfo {author} {\bibfnamefont {Y.}~\bibnamefont
  {Uemura}},\ }\href@noop {} {\bibfield  {journal} {\bibinfo  {journal}
  {Journal of the Physical Society of Japan}\ }\textbf {\bibinfo {volume}
  {37}},\ \bibinfo {pages} {1325} (\bibinfo {year} {1974})}\BibitemShut
  {NoStop}%
\bibitem [{\citenamefont {Bychkov}\ and\ \citenamefont
  {Rashba}(1984)}]{RashbaSOI}%
  \BibitemOpen
  \bibfield  {author} {\bibinfo {author} {\bibfnamefont {Y.~A.}\ \bibnamefont
  {Bychkov}}\ and\ \bibinfo {author} {\bibfnamefont {E.~I.}\ \bibnamefont
  {Rashba}},\ }\href@noop {} {\bibfield  {journal} {\bibinfo  {journal}
  {Journal of physics C: Solid state physics}\ }\textbf {\bibinfo {volume}
  {17}},\ \bibinfo {pages} {6039} (\bibinfo {year} {1984})}\BibitemShut
  {NoStop}%
\bibitem [{\citenamefont {Schierholz}(2005)}]{Rashba4}%
  \BibitemOpen
  \bibfield  {author} {\bibinfo {author} {\bibfnamefont {C.}~\bibnamefont
  {Schierholz}},\ }\href@noop {} {\emph {\bibinfo {title} {Rashba Spin-Orbit
  Interaction in Low and High Magnetic Fields}}}\ (\bibinfo  {publisher}
  {Cuvillier Verlag},\ \bibinfo {year} {2005})\BibitemShut {NoStop}%
\bibitem [{\citenamefont {Str\"om}\ \emph {et~al.}(2010)\citenamefont
  {Str\"om}, \citenamefont {Johannesson},\ and\ \citenamefont
  {Japaridze}}]{Rashba1}%
  \BibitemOpen
  \bibfield  {author} {\bibinfo {author} {\bibfnamefont {A.}~\bibnamefont
  {Str\"om}}, \bibinfo {author} {\bibfnamefont {H.}~\bibnamefont
  {Johannesson}}, \ and\ \bibinfo {author} {\bibfnamefont {G.~I.}\ \bibnamefont
  {Japaridze}},\ }\href {\doibase 10.1103/PhysRevLett.104.256804} {\bibfield
  {journal} {\bibinfo  {journal} {Phys. Rev. Lett.}\ }\textbf {\bibinfo
  {volume} {104}},\ \bibinfo {pages} {256804} (\bibinfo {year}
  {2010})}\BibitemShut {NoStop}%
\bibitem [{\citenamefont {Cr\'epin}\ \emph {et~al.}(2012)\citenamefont
  {Cr\'epin}, \citenamefont {Budich}, \citenamefont {Dolcini}, \citenamefont
  {Recher},\ and\ \citenamefont {Trauzettel}}]{Rashba2}%
  \BibitemOpen
  \bibfield  {author} {\bibinfo {author} {\bibfnamefont {F.~m.~c.}\
  \bibnamefont {Cr\'epin}}, \bibinfo {author} {\bibfnamefont {J.~C.}\
  \bibnamefont {Budich}}, \bibinfo {author} {\bibfnamefont {F.}~\bibnamefont
  {Dolcini}}, \bibinfo {author} {\bibfnamefont {P.}~\bibnamefont {Recher}}, \
  and\ \bibinfo {author} {\bibfnamefont {B.}~\bibnamefont {Trauzettel}},\
  }\href {\doibase 10.1103/PhysRevB.86.121106} {\bibfield  {journal} {\bibinfo
  {journal} {Phys. Rev. B}\ }\textbf {\bibinfo {volume} {86}},\ \bibinfo
  {pages} {121106} (\bibinfo {year} {2012})}\BibitemShut {NoStop}%
\bibitem [{\citenamefont {Geissler}\ \emph {et~al.}(2014)\citenamefont
  {Geissler}, \citenamefont {Cr\'epin},\ and\ \citenamefont
  {Trauzettel}}]{Rashba3}%
  \BibitemOpen
  \bibfield  {author} {\bibinfo {author} {\bibfnamefont {F.}~\bibnamefont
  {Geissler}}, \bibinfo {author} {\bibfnamefont {F.~m.~c.}\ \bibnamefont
  {Cr\'epin}}, \ and\ \bibinfo {author} {\bibfnamefont {B.}~\bibnamefont
  {Trauzettel}},\ }\href {\doibase 10.1103/PhysRevB.89.235136} {\bibfield
  {journal} {\bibinfo  {journal} {Phys. Rev. B}\ }\textbf {\bibinfo {volume}
  {89}},\ \bibinfo {pages} {235136} (\bibinfo {year} {2014})}\BibitemShut
  {NoStop}%
\bibitem [{\citenamefont {Manchon}\ \emph {et~al.}(2015)\citenamefont
  {Manchon}, \citenamefont {Koo}, \citenamefont {Nitta}, \citenamefont
  {Frolov},\ and\ \citenamefont {Duine}}]{RashbaSOI3}%
  \BibitemOpen
  \bibfield  {author} {\bibinfo {author} {\bibfnamefont {A.}~\bibnamefont
  {Manchon}}, \bibinfo {author} {\bibfnamefont {H.~C.}\ \bibnamefont {Koo}},
  \bibinfo {author} {\bibfnamefont {J.}~\bibnamefont {Nitta}}, \bibinfo
  {author} {\bibfnamefont {S.~M.}\ \bibnamefont {Frolov}}, \ and\ \bibinfo
  {author} {\bibfnamefont {R.~A.}\ \bibnamefont {Duine}},\ }\href@noop {}
  {\bibfield  {journal} {\bibinfo  {journal} {Nature materials}\ }\textbf
  {\bibinfo {volume} {14 9}},\ \bibinfo {pages} {871} (\bibinfo {year}
  {2015})}\BibitemShut {NoStop}%
\bibitem [{\citenamefont {\ifmmode \check{Z}\else
  \v{Z}\fi{}uti\ifmmode~\acute{c}\else \'{c}\fi{}}\ \emph
  {et~al.}(2004)\citenamefont {\ifmmode \check{Z}\else
  \v{Z}\fi{}uti\ifmmode~\acute{c}\else \'{c}\fi{}}, \citenamefont {Fabian},\
  and\ \citenamefont {Das~Sarma}}]{Spintronics1}%
  \BibitemOpen
  \bibfield  {author} {\bibinfo {author} {\bibfnamefont {I.}~\bibnamefont
  {\ifmmode \check{Z}\else \v{Z}\fi{}uti\ifmmode~\acute{c}\else \'{c}\fi{}}},
  \bibinfo {author} {\bibfnamefont {J.}~\bibnamefont {Fabian}}, \ and\ \bibinfo
  {author} {\bibfnamefont {S.}~\bibnamefont {Das~Sarma}},\ }\href {\doibase
  10.1103/RevModPhys.76.323} {\bibfield  {journal} {\bibinfo  {journal} {Rev.
  Mod. Phys.}\ }\textbf {\bibinfo {volume} {76}},\ \bibinfo {pages} {323}
  (\bibinfo {year} {2004})}\BibitemShut {NoStop}%
\bibitem [{\citenamefont {Fiederling}\ \emph {et~al.}(1999)\citenamefont
  {Fiederling}, \citenamefont {Keim}, \citenamefont {Reuscher}, \citenamefont
  {Ossau}, \citenamefont {Schmidt}, \citenamefont {Waag},\ and\ \citenamefont
  {Molenkamp}}]{Spintronics3}%
  \BibitemOpen
  \bibfield  {author} {\bibinfo {author} {\bibfnamefont {R.}~\bibnamefont
  {Fiederling}}, \bibinfo {author} {\bibfnamefont {M.}~\bibnamefont {Keim}},
  \bibinfo {author} {\bibfnamefont {G.}~\bibnamefont {Reuscher}}, \bibinfo
  {author} {\bibfnamefont {W.}~\bibnamefont {Ossau}}, \bibinfo {author}
  {\bibfnamefont {G.}~\bibnamefont {Schmidt}}, \bibinfo {author} {\bibfnamefont
  {A.}~\bibnamefont {Waag}}, \ and\ \bibinfo {author} {\bibfnamefont {L.~W.}\
  \bibnamefont {Molenkamp}},\ }\href {\doibase 10.1038/45502} {\bibfield
  {journal} {\bibinfo  {journal} {Nature}\ }\textbf {\bibinfo {volume} {402}},\
  \bibinfo {pages} {787} (\bibinfo {year} {1999})}\BibitemShut {NoStop}%
\bibitem [{\citenamefont {Br{\"u}ne}\ \emph {et~al.}(2012)\citenamefont
  {Br{\"u}ne}, \citenamefont {Roth}, \citenamefont {Buhmann}, \citenamefont
  {Hankiewicz}, \citenamefont {Molenkamp}, \citenamefont {Maciejko},
  \citenamefont {Qi},\ and\ \citenamefont {Zhang}}]{SHE}%
  \BibitemOpen
  \bibfield  {author} {\bibinfo {author} {\bibfnamefont {C.}~\bibnamefont
  {Br{\"u}ne}}, \bibinfo {author} {\bibfnamefont {A.}~\bibnamefont {Roth}},
  \bibinfo {author} {\bibfnamefont {H.}~\bibnamefont {Buhmann}}, \bibinfo
  {author} {\bibfnamefont {E.~M.}\ \bibnamefont {Hankiewicz}}, \bibinfo
  {author} {\bibfnamefont {L.~W.}\ \bibnamefont {Molenkamp}}, \bibinfo {author}
  {\bibfnamefont {J.}~\bibnamefont {Maciejko}}, \bibinfo {author}
  {\bibfnamefont {X.-L.}\ \bibnamefont {Qi}}, \ and\ \bibinfo {author}
  {\bibfnamefont {S.-C.}\ \bibnamefont {Zhang}},\ }\href
  {https://doi.org/10.1038/nphys2322} {\bibfield  {journal} {\bibinfo
  {journal} {Nature Physics}\ }\textbf {\bibinfo {volume} {8}},\ \bibinfo
  {pages} {485 EP } (\bibinfo {year} {2012})},\ \bibinfo {note}
  {article}\BibitemShut {NoStop}%
\bibitem [{\citenamefont {Jusserand}\ \emph {et~al.}(1995)\citenamefont
  {Jusserand}, \citenamefont {Richards}, \citenamefont {Allan}, \citenamefont
  {Priester},\ and\ \citenamefont {Etienne}}]{RashbaStrength2}%
  \BibitemOpen
  \bibfield  {author} {\bibinfo {author} {\bibfnamefont {B.}~\bibnamefont
  {Jusserand}}, \bibinfo {author} {\bibfnamefont {D.}~\bibnamefont {Richards}},
  \bibinfo {author} {\bibfnamefont {G.}~\bibnamefont {Allan}}, \bibinfo
  {author} {\bibfnamefont {C.}~\bibnamefont {Priester}}, \ and\ \bibinfo
  {author} {\bibfnamefont {B.}~\bibnamefont {Etienne}},\ }\href@noop {}
  {\bibfield  {journal} {\bibinfo  {journal} {Physical Review B}\ }\textbf
  {\bibinfo {volume} {51}},\ \bibinfo {pages} {4707} (\bibinfo {year}
  {1995})}\BibitemShut {NoStop}%
\bibitem [{\citenamefont {Qiu}\ \emph {et~al.}(2004)\citenamefont {Qiu},
  \citenamefont {Gui}, \citenamefont {Lin}, \citenamefont {Dai}, \citenamefont
  {Chu}, \citenamefont {Tang}, \citenamefont {Lu},\ and\ \citenamefont
  {Shen}}]{WL4}%
  \BibitemOpen
  \bibfield  {author} {\bibinfo {author} {\bibfnamefont {Z.~J.}\ \bibnamefont
  {Qiu}}, \bibinfo {author} {\bibfnamefont {Y.~S.}\ \bibnamefont {Gui}},
  \bibinfo {author} {\bibfnamefont {T.}~\bibnamefont {Lin}}, \bibinfo {author}
  {\bibfnamefont {N.}~\bibnamefont {Dai}}, \bibinfo {author} {\bibfnamefont
  {J.~H.}\ \bibnamefont {Chu}}, \bibinfo {author} {\bibfnamefont
  {N.}~\bibnamefont {Tang}}, \bibinfo {author} {\bibfnamefont {J.}~\bibnamefont
  {Lu}}, \ and\ \bibinfo {author} {\bibfnamefont {B.}~\bibnamefont {Shen}},\
  }\href {\doibase 10.1103/PhysRevB.69.125335} {\bibfield  {journal} {\bibinfo
  {journal} {Phys. Rev. B}\ }\textbf {\bibinfo {volume} {69}},\ \bibinfo
  {pages} {125335} (\bibinfo {year} {2004})}\BibitemShut {NoStop}%
\bibitem [{\citenamefont {Tikhonenko}\ \emph {et~al.}(2008)\citenamefont
  {Tikhonenko}, \citenamefont {Horsell}, \citenamefont {Gorbachev},\ and\
  \citenamefont {Savchenko}}]{WL5}%
  \BibitemOpen
  \bibfield  {author} {\bibinfo {author} {\bibfnamefont {F.~V.}\ \bibnamefont
  {Tikhonenko}}, \bibinfo {author} {\bibfnamefont {D.~W.}\ \bibnamefont
  {Horsell}}, \bibinfo {author} {\bibfnamefont {R.~V.}\ \bibnamefont
  {Gorbachev}}, \ and\ \bibinfo {author} {\bibfnamefont {A.~K.}\ \bibnamefont
  {Savchenko}},\ }\href {\doibase 10.1103/PhysRevLett.100.056802} {\bibfield
  {journal} {\bibinfo  {journal} {Phys. Rev. Lett.}\ }\textbf {\bibinfo
  {volume} {100}},\ \bibinfo {pages} {056802} (\bibinfo {year}
  {2008})}\BibitemShut {NoStop}%
\bibitem [{\citenamefont {Studer}\ \emph {et~al.}(2009)\citenamefont {Studer},
  \citenamefont {Salis}, \citenamefont {Ensslin}, \citenamefont {Driscoll},\
  and\ \citenamefont {Gossard}}]{WL6}%
  \BibitemOpen
  \bibfield  {author} {\bibinfo {author} {\bibfnamefont {M.}~\bibnamefont
  {Studer}}, \bibinfo {author} {\bibfnamefont {G.}~\bibnamefont {Salis}},
  \bibinfo {author} {\bibfnamefont {K.}~\bibnamefont {Ensslin}}, \bibinfo
  {author} {\bibfnamefont {D.~C.}\ \bibnamefont {Driscoll}}, \ and\ \bibinfo
  {author} {\bibfnamefont {A.~C.}\ \bibnamefont {Gossard}},\ }\href {\doibase
  10.1103/PhysRevLett.103.027201} {\bibfield  {journal} {\bibinfo  {journal}
  {Phys. Rev. Lett.}\ }\textbf {\bibinfo {volume} {103}},\ \bibinfo {pages}
  {027201} (\bibinfo {year} {2009})}\BibitemShut {NoStop}%
\bibitem [{\citenamefont {Wang}\ \emph {et~al.}(2018)\citenamefont {Wang},
  \citenamefont {Yin}, \citenamefont {Khan}, \citenamefont {Muhtadi},
  \citenamefont {Asif}, \citenamefont {Choi},\ and\ \citenamefont
  {Datta}}]{WL3}%
  \BibitemOpen
  \bibfield  {author} {\bibinfo {author} {\bibfnamefont {L.}~\bibnamefont
  {Wang}}, \bibinfo {author} {\bibfnamefont {M.}~\bibnamefont {Yin}}, \bibinfo
  {author} {\bibfnamefont {A.}~\bibnamefont {Khan}}, \bibinfo {author}
  {\bibfnamefont {S.}~\bibnamefont {Muhtadi}}, \bibinfo {author} {\bibfnamefont
  {F.}~\bibnamefont {Asif}}, \bibinfo {author} {\bibfnamefont {E.~S.}\
  \bibnamefont {Choi}}, \ and\ \bibinfo {author} {\bibfnamefont
  {T.}~\bibnamefont {Datta}},\ }\href {\doibase
  10.1103/PhysRevApplied.9.024006} {\bibfield  {journal} {\bibinfo  {journal}
  {Phys. Rev. Applied}\ }\textbf {\bibinfo {volume} {9}},\ \bibinfo {pages}
  {024006} (\bibinfo {year} {2018})}\BibitemShut {NoStop}%
\bibitem [{\citenamefont {Luo}\ \emph {et~al.}(1990)\citenamefont {Luo},
  \citenamefont {Munekata}, \citenamefont {Fang},\ and\ \citenamefont
  {Stiles}}]{WL8}%
  \BibitemOpen
  \bibfield  {author} {\bibinfo {author} {\bibfnamefont {J.}~\bibnamefont
  {Luo}}, \bibinfo {author} {\bibfnamefont {H.}~\bibnamefont {Munekata}},
  \bibinfo {author} {\bibfnamefont {F.~F.}\ \bibnamefont {Fang}}, \ and\
  \bibinfo {author} {\bibfnamefont {P.~J.}\ \bibnamefont {Stiles}},\ }\href
  {\doibase 10.1103/PhysRevB.41.7685} {\bibfield  {journal} {\bibinfo
  {journal} {Phys. Rev. B}\ }\textbf {\bibinfo {volume} {41}},\ \bibinfo
  {pages} {7685} (\bibinfo {year} {1990})}\BibitemShut {NoStop}%
\bibitem [{\citenamefont {Nitta}\ \emph {et~al.}(1997)\citenamefont {Nitta},
  \citenamefont {Akazaki}, \citenamefont {Takayanagi},\ and\ \citenamefont
  {Enoki}}]{RashbaStrength3}%
  \BibitemOpen
  \bibfield  {author} {\bibinfo {author} {\bibfnamefont {J.}~\bibnamefont
  {Nitta}}, \bibinfo {author} {\bibfnamefont {T.}~\bibnamefont {Akazaki}},
  \bibinfo {author} {\bibfnamefont {H.}~\bibnamefont {Takayanagi}}, \ and\
  \bibinfo {author} {\bibfnamefont {T.}~\bibnamefont {Enoki}},\ }\href@noop {}
  {\bibfield  {journal} {\bibinfo  {journal} {Physical Review Letters}\
  }\textbf {\bibinfo {volume} {78}},\ \bibinfo {pages} {1335} (\bibinfo {year}
  {1997})}\BibitemShut {NoStop}%
\bibitem [{\citenamefont {Engels}\ \emph {et~al.}(1997)\citenamefont {Engels},
  \citenamefont {Lange}, \citenamefont {Sch\"apers},\ and\ \citenamefont
  {L\"uth}}]{RashbaStrength4}%
  \BibitemOpen
  \bibfield  {author} {\bibinfo {author} {\bibfnamefont {G.}~\bibnamefont
  {Engels}}, \bibinfo {author} {\bibfnamefont {J.}~\bibnamefont {Lange}},
  \bibinfo {author} {\bibfnamefont {T.}~\bibnamefont {Sch\"apers}}, \ and\
  \bibinfo {author} {\bibfnamefont {H.}~\bibnamefont {L\"uth}},\ }\href
  {\doibase 10.1103/PhysRevB.55.R1958} {\bibfield  {journal} {\bibinfo
  {journal} {Phys. Rev. B}\ }\textbf {\bibinfo {volume} {55}},\ \bibinfo
  {pages} {R1958} (\bibinfo {year} {1997})}\BibitemShut {NoStop}%
\bibitem [{\citenamefont {Heida}\ \emph {et~al.}(1998)\citenamefont {Heida},
  \citenamefont {Van~Wees}, \citenamefont {Kuipers}, \citenamefont {Klapwijk},\
  and\ \citenamefont {Borghs}}]{RashbaStrength5}%
  \BibitemOpen
  \bibfield  {author} {\bibinfo {author} {\bibfnamefont {J.}~\bibnamefont
  {Heida}}, \bibinfo {author} {\bibfnamefont {B.}~\bibnamefont {Van~Wees}},
  \bibinfo {author} {\bibfnamefont {J.}~\bibnamefont {Kuipers}}, \bibinfo
  {author} {\bibfnamefont {T.}~\bibnamefont {Klapwijk}}, \ and\ \bibinfo
  {author} {\bibfnamefont {G.}~\bibnamefont {Borghs}},\ }\href@noop {}
  {\bibfield  {journal} {\bibinfo  {journal} {Physical Review B}\ }\textbf
  {\bibinfo {volume} {57}},\ \bibinfo {pages} {11911} (\bibinfo {year}
  {1998})}\BibitemShut {NoStop}%
\bibitem [{\citenamefont {Hu}\ \emph {et~al.}(1999)\citenamefont {Hu},
  \citenamefont {Nitta}, \citenamefont {Akazaki}, \citenamefont {Takayanagi},
  \citenamefont {Osaka}, \citenamefont {Pfeffer},\ and\ \citenamefont
  {Zawadzki}}]{RashbaStrength6}%
  \BibitemOpen
  \bibfield  {author} {\bibinfo {author} {\bibfnamefont {C.-M.}\ \bibnamefont
  {Hu}}, \bibinfo {author} {\bibfnamefont {J.}~\bibnamefont {Nitta}}, \bibinfo
  {author} {\bibfnamefont {T.}~\bibnamefont {Akazaki}}, \bibinfo {author}
  {\bibfnamefont {H.}~\bibnamefont {Takayanagi}}, \bibinfo {author}
  {\bibfnamefont {J.}~\bibnamefont {Osaka}}, \bibinfo {author} {\bibfnamefont
  {P.}~\bibnamefont {Pfeffer}}, \ and\ \bibinfo {author} {\bibfnamefont
  {W.}~\bibnamefont {Zawadzki}},\ }\href@noop {} {\bibfield  {journal}
  {\bibinfo  {journal} {Physical Review B}\ }\textbf {\bibinfo {volume} {60}},\
  \bibinfo {pages} {7736} (\bibinfo {year} {1999})}\BibitemShut {NoStop}%
\bibitem [{\citenamefont {Brosig}\ \emph {et~al.}(1999)\citenamefont {Brosig},
  \citenamefont {Ensslin}, \citenamefont {Warburton}, \citenamefont {Nguyen},
  \citenamefont {Brar}, \citenamefont {Thomas},\ and\ \citenamefont
  {Kroemer}}]{RashbaAbsent}%
  \BibitemOpen
  \bibfield  {author} {\bibinfo {author} {\bibfnamefont {S.}~\bibnamefont
  {Brosig}}, \bibinfo {author} {\bibfnamefont {K.}~\bibnamefont {Ensslin}},
  \bibinfo {author} {\bibfnamefont {R.}~\bibnamefont {Warburton}}, \bibinfo
  {author} {\bibfnamefont {C.}~\bibnamefont {Nguyen}}, \bibinfo {author}
  {\bibfnamefont {B.}~\bibnamefont {Brar}}, \bibinfo {author} {\bibfnamefont
  {M.}~\bibnamefont {Thomas}}, \ and\ \bibinfo {author} {\bibfnamefont
  {H.}~\bibnamefont {Kroemer}},\ }\href@noop {} {\bibfield  {journal} {\bibinfo
   {journal} {Physical Review B}\ }\textbf {\bibinfo {volume} {60}},\ \bibinfo
  {pages} {R13989} (\bibinfo {year} {1999})}\BibitemShut {NoStop}%
\bibitem [{\citenamefont {Papadakis}\ \emph {et~al.}(1999)\citenamefont
  {Papadakis}, \citenamefont {De~Poortere}, \citenamefont {Manoharan},
  \citenamefont {Shayegan},\ and\ \citenamefont {Winkler}}]{RashbaStrong}%
  \BibitemOpen
  \bibfield  {author} {\bibinfo {author} {\bibfnamefont {S.}~\bibnamefont
  {Papadakis}}, \bibinfo {author} {\bibfnamefont {E.}~\bibnamefont
  {De~Poortere}}, \bibinfo {author} {\bibfnamefont {H.}~\bibnamefont
  {Manoharan}}, \bibinfo {author} {\bibfnamefont {M.}~\bibnamefont {Shayegan}},
  \ and\ \bibinfo {author} {\bibfnamefont {R.}~\bibnamefont {Winkler}},\
  }\href@noop {} {\bibfield  {journal} {\bibinfo  {journal} {Science}\ }\textbf
  {\bibinfo {volume} {283}},\ \bibinfo {pages} {2056} (\bibinfo {year}
  {1999})}\BibitemShut {NoStop}%
\bibitem [{\citenamefont {Grundler}(2000)}]{RashbaGate}%
  \BibitemOpen
  \bibfield  {author} {\bibinfo {author} {\bibfnamefont {D.}~\bibnamefont
  {Grundler}},\ }\href {\doibase 10.1103/PhysRevLett.84.6074} {\bibfield
  {journal} {\bibinfo  {journal} {Phys. Rev. Lett.}\ }\textbf {\bibinfo
  {volume} {84}},\ \bibinfo {pages} {6074} (\bibinfo {year}
  {2000})}\BibitemShut {NoStop}%
\bibitem [{\citenamefont {Koga}\ \emph {et~al.}(2002)\citenamefont {Koga},
  \citenamefont {Nitta}, \citenamefont {Akazaki},\ and\ \citenamefont
  {Takayanagi}}]{Nitta}%
  \BibitemOpen
  \bibfield  {author} {\bibinfo {author} {\bibfnamefont {T.}~\bibnamefont
  {Koga}}, \bibinfo {author} {\bibfnamefont {J.}~\bibnamefont {Nitta}},
  \bibinfo {author} {\bibfnamefont {T.}~\bibnamefont {Akazaki}}, \ and\
  \bibinfo {author} {\bibfnamefont {H.}~\bibnamefont {Takayanagi}},\ }\href
  {\doibase 10.1103/PhysRevLett.89.046801} {\bibfield  {journal} {\bibinfo
  {journal} {Phys. Rev. Lett.}\ }\textbf {\bibinfo {volume} {89}},\ \bibinfo
  {pages} {046801} (\bibinfo {year} {2002})}\BibitemShut {NoStop}%
\bibitem [{\citenamefont {Shojaei}\ \emph {et~al.}(2016)\citenamefont
  {Shojaei}, \citenamefont {O'Malley}, \citenamefont {Shabani}, \citenamefont
  {Roushan}, \citenamefont {Schultz}, \citenamefont {Lutchyn}, \citenamefont
  {Nayak}, \citenamefont {Martinis},\ and\ \citenamefont
  {Palmstr{\o}m}}]{BeatSdH3}%
  \BibitemOpen
  \bibfield  {author} {\bibinfo {author} {\bibfnamefont {B.}~\bibnamefont
  {Shojaei}}, \bibinfo {author} {\bibfnamefont {P.}~\bibnamefont {O'Malley}},
  \bibinfo {author} {\bibfnamefont {J.}~\bibnamefont {Shabani}}, \bibinfo
  {author} {\bibfnamefont {P.}~\bibnamefont {Roushan}}, \bibinfo {author}
  {\bibfnamefont {B.}~\bibnamefont {Schultz}}, \bibinfo {author} {\bibfnamefont
  {R.}~\bibnamefont {Lutchyn}}, \bibinfo {author} {\bibfnamefont
  {C.}~\bibnamefont {Nayak}}, \bibinfo {author} {\bibfnamefont
  {J.}~\bibnamefont {Martinis}}, \ and\ \bibinfo {author} {\bibfnamefont
  {C.}~\bibnamefont {Palmstr{\o}m}},\ }\href@noop {} {\bibfield  {journal}
  {\bibinfo  {journal} {Physical Review B}\ }\textbf {\bibinfo {volume} {93}},\
  \bibinfo {pages} {075302} (\bibinfo {year} {2016})}\BibitemShut {NoStop}%
\bibitem [{\citenamefont {Beukman}\ \emph {et~al.}(2017)\citenamefont
  {Beukman}, \citenamefont {de~Vries}, \citenamefont {van Veen}, \citenamefont
  {Skolasinski}, \citenamefont {Wimmer}, \citenamefont {Qu}, \citenamefont
  {de~Vries}, \citenamefont {Nguyen}, \citenamefont {Yi}, \citenamefont
  {Kiselev}, \citenamefont {Sokolich}, \citenamefont {Manfra}, \citenamefont
  {Nichele}, \citenamefont {Marcus},\ and\ \citenamefont
  {Kouwenhoven}}]{Beukman_BeatPattern}%
  \BibitemOpen
  \bibfield  {author} {\bibinfo {author} {\bibfnamefont {A.~J.~A.}\
  \bibnamefont {Beukman}}, \bibinfo {author} {\bibfnamefont {F.~K.}\
  \bibnamefont {de~Vries}}, \bibinfo {author} {\bibfnamefont {J.}~\bibnamefont
  {van Veen}}, \bibinfo {author} {\bibfnamefont {R.}~\bibnamefont
  {Skolasinski}}, \bibinfo {author} {\bibfnamefont {M.}~\bibnamefont {Wimmer}},
  \bibinfo {author} {\bibfnamefont {F.}~\bibnamefont {Qu}}, \bibinfo {author}
  {\bibfnamefont {D.~T.}\ \bibnamefont {de~Vries}}, \bibinfo {author}
  {\bibfnamefont {B.-M.}\ \bibnamefont {Nguyen}}, \bibinfo {author}
  {\bibfnamefont {W.}~\bibnamefont {Yi}}, \bibinfo {author} {\bibfnamefont
  {A.~A.}\ \bibnamefont {Kiselev}}, \bibinfo {author} {\bibfnamefont
  {M.}~\bibnamefont {Sokolich}}, \bibinfo {author} {\bibfnamefont {M.~J.}\
  \bibnamefont {Manfra}}, \bibinfo {author} {\bibfnamefont {F.}~\bibnamefont
  {Nichele}}, \bibinfo {author} {\bibfnamefont {C.~M.}\ \bibnamefont {Marcus}},
  \ and\ \bibinfo {author} {\bibfnamefont {L.~P.}\ \bibnamefont
  {Kouwenhoven}},\ }\href {\doibase 10.1103/PhysRevB.96.241401} {\bibfield
  {journal} {\bibinfo  {journal} {Phys. Rev. B}\ }\textbf {\bibinfo {volume}
  {96}},\ \bibinfo {pages} {241401} (\bibinfo {year} {2017})}\BibitemShut
  {NoStop}%
\bibitem [{\citenamefont {Herzog}\ \emph {et~al.}(2017)\citenamefont {Herzog},
  \citenamefont {Hardtdegen}, \citenamefont {Sch{\"a}pers}, \citenamefont
  {Grundler},\ and\ \citenamefont {Wilde}}]{BeatSdH2}%
  \BibitemOpen
  \bibfield  {author} {\bibinfo {author} {\bibfnamefont {F.}~\bibnamefont
  {Herzog}}, \bibinfo {author} {\bibfnamefont {H.}~\bibnamefont {Hardtdegen}},
  \bibinfo {author} {\bibfnamefont {T.}~\bibnamefont {Sch{\"a}pers}}, \bibinfo
  {author} {\bibfnamefont {D.}~\bibnamefont {Grundler}}, \ and\ \bibinfo
  {author} {\bibfnamefont {M.}~\bibnamefont {Wilde}},\ }\href@noop {}
  {\bibfield  {journal} {\bibinfo  {journal} {New Journal of Physics}\ }\textbf
  {\bibinfo {volume} {19}},\ \bibinfo {pages} {103012} (\bibinfo {year}
  {2017})}\BibitemShut {NoStop}%
\bibitem [{\citenamefont {Nichele}\ \emph {et~al.}(2017)\citenamefont
  {Nichele}, \citenamefont {Kjaergaard}, \citenamefont {Suominen},
  \citenamefont {Skolasinski}, \citenamefont {Wimmer}, \citenamefont {Nguyen},
  \citenamefont {Kiselev}, \citenamefont {Yi}, \citenamefont {Sokolich},
  \citenamefont {Manfra}, \citenamefont {Qu}, \citenamefont {Beukman},
  \citenamefont {Kouwenhoven},\ and\ \citenamefont
  {Marcus}}]{Nichele_BeatingPattern}%
  \BibitemOpen
  \bibfield  {author} {\bibinfo {author} {\bibfnamefont {F.}~\bibnamefont
  {Nichele}}, \bibinfo {author} {\bibfnamefont {M.}~\bibnamefont {Kjaergaard}},
  \bibinfo {author} {\bibfnamefont {H.~J.}\ \bibnamefont {Suominen}}, \bibinfo
  {author} {\bibfnamefont {R.}~\bibnamefont {Skolasinski}}, \bibinfo {author}
  {\bibfnamefont {M.}~\bibnamefont {Wimmer}}, \bibinfo {author} {\bibfnamefont
  {B.-M.}\ \bibnamefont {Nguyen}}, \bibinfo {author} {\bibfnamefont {A.~A.}\
  \bibnamefont {Kiselev}}, \bibinfo {author} {\bibfnamefont {W.}~\bibnamefont
  {Yi}}, \bibinfo {author} {\bibfnamefont {M.}~\bibnamefont {Sokolich}},
  \bibinfo {author} {\bibfnamefont {M.~J.}\ \bibnamefont {Manfra}}, \bibinfo
  {author} {\bibfnamefont {F.}~\bibnamefont {Qu}}, \bibinfo {author}
  {\bibfnamefont {A.~J.~A.}\ \bibnamefont {Beukman}}, \bibinfo {author}
  {\bibfnamefont {L.~P.}\ \bibnamefont {Kouwenhoven}}, \ and\ \bibinfo {author}
  {\bibfnamefont {C.~M.}\ \bibnamefont {Marcus}},\ }\href {\doibase
  10.1103/PhysRevLett.118.016801} {\bibfield  {journal} {\bibinfo  {journal}
  {Phys. Rev. Lett.}\ }\textbf {\bibinfo {volume} {118}},\ \bibinfo {pages}
  {016801} (\bibinfo {year} {2017})}\BibitemShut {NoStop}%
\bibitem [{\citenamefont {Herling}\ \emph {et~al.}(2017)\citenamefont
  {Herling}, \citenamefont {Morrison}, \citenamefont {Knox}, \citenamefont
  {Zhang}, \citenamefont {Newell}, \citenamefont {Myronov}, \citenamefont
  {Linfield},\ and\ \citenamefont {Marrows}}]{WL2}%
  \BibitemOpen
  \bibfield  {author} {\bibinfo {author} {\bibfnamefont {F.}~\bibnamefont
  {Herling}}, \bibinfo {author} {\bibfnamefont {C.}~\bibnamefont {Morrison}},
  \bibinfo {author} {\bibfnamefont {C.~S.}\ \bibnamefont {Knox}}, \bibinfo
  {author} {\bibfnamefont {S.}~\bibnamefont {Zhang}}, \bibinfo {author}
  {\bibfnamefont {O.}~\bibnamefont {Newell}}, \bibinfo {author} {\bibfnamefont
  {M.}~\bibnamefont {Myronov}}, \bibinfo {author} {\bibfnamefont {E.~H.}\
  \bibnamefont {Linfield}}, \ and\ \bibinfo {author} {\bibfnamefont {C.~H.}\
  \bibnamefont {Marrows}},\ }\href {\doibase 10.1103/PhysRevB.95.155307}
  {\bibfield  {journal} {\bibinfo  {journal} {Phys. Rev. B}\ }\textbf {\bibinfo
  {volume} {95}},\ \bibinfo {pages} {155307} (\bibinfo {year}
  {2017})}\BibitemShut {NoStop}%
\bibitem [{\citenamefont {Assaf}\ \emph {et~al.}(2013)\citenamefont {Assaf},
  \citenamefont {Cardinal}, \citenamefont {Wei}, \citenamefont {Katmis},
  \citenamefont {Moodera},\ and\ \citenamefont {Heiman}}]{WL7}%
  \BibitemOpen
  \bibfield  {author} {\bibinfo {author} {\bibfnamefont {B.~A.}\ \bibnamefont
  {Assaf}}, \bibinfo {author} {\bibfnamefont {T.}~\bibnamefont {Cardinal}},
  \bibinfo {author} {\bibfnamefont {P.}~\bibnamefont {Wei}}, \bibinfo {author}
  {\bibfnamefont {F.}~\bibnamefont {Katmis}}, \bibinfo {author} {\bibfnamefont
  {J.~S.}\ \bibnamefont {Moodera}}, \ and\ \bibinfo {author} {\bibfnamefont
  {D.}~\bibnamefont {Heiman}},\ }\href {\doibase 10.1063/1.4773207} {\bibfield
  {journal} {\bibinfo  {journal} {Applied Physics Letters}\ }\textbf {\bibinfo
  {volume} {102}},\ \bibinfo {pages} {012102} (\bibinfo {year} {2013})},\
  \Eprint {http://arxiv.org/abs/https://doi.org/10.1063/1.4773207}
  {https://doi.org/10.1063/1.4773207} \BibitemShut {NoStop}%
\bibitem [{\citenamefont {Meng}\ \emph {et~al.}(2018)\citenamefont {Meng},
  \citenamefont {Huang}, \citenamefont {Tan}, \citenamefont {Wu}, \citenamefont
  {Jing}, \citenamefont {Peng},\ and\ \citenamefont {Xu}}]{WL1}%
  \BibitemOpen
  \bibfield  {author} {\bibinfo {author} {\bibfnamefont {M.}~\bibnamefont
  {Meng}}, \bibinfo {author} {\bibfnamefont {S.}~\bibnamefont {Huang}},
  \bibinfo {author} {\bibfnamefont {C.}~\bibnamefont {Tan}}, \bibinfo {author}
  {\bibfnamefont {J.}~\bibnamefont {Wu}}, \bibinfo {author} {\bibfnamefont
  {Y.}~\bibnamefont {Jing}}, \bibinfo {author} {\bibfnamefont {H.}~\bibnamefont
  {Peng}}, \ and\ \bibinfo {author} {\bibfnamefont {H.~Q.}\ \bibnamefont
  {Xu}},\ }\href {\doibase 10.1039/C7NR08874D} {\bibfield  {journal} {\bibinfo
  {journal} {Nanoscale}\ }\textbf {\bibinfo {volume} {10}},\ \bibinfo {pages}
  {2704} (\bibinfo {year} {2018})}\BibitemShut {NoStop}%
\bibitem [{\citenamefont {Hikami}\ \emph {et~al.}(1980)\citenamefont {Hikami},
  \citenamefont {Larkin},\ and\ \citenamefont {Nagaoka}}]{HLN}%
  \BibitemOpen
  \bibfield  {author} {\bibinfo {author} {\bibfnamefont {S.}~\bibnamefont
  {Hikami}}, \bibinfo {author} {\bibfnamefont {A.~I.}\ \bibnamefont {Larkin}},
  \ and\ \bibinfo {author} {\bibfnamefont {Y.}~\bibnamefont {Nagaoka}},\ }\href
  {\doibase 10.1143/PTP.63.707} {\bibfield  {journal} {\bibinfo  {journal}
  {Progress of Theoretical Physics}\ }\textbf {\bibinfo {volume} {63}},\
  \bibinfo {pages} {707} (\bibinfo {year} {1980})},\ \Eprint
  {http://arxiv.org/abs/http://oup.prod.sis.lan/ptp/article-pdf/63/2/707/5336056/63-2-707.pdf}
  {http://oup.prod.sis.lan/ptp/article-pdf/63/2/707/5336056/63-2-707.pdf}
  \BibitemShut {NoStop}%
\bibitem [{\citenamefont {Iordanskii}\ \emph {et~al.}(1994)\citenamefont
  {Iordanskii}, \citenamefont {Lyanda-Geller},\ and\ \citenamefont
  {Pikus}}]{WL-HLN}%
  \BibitemOpen
  \bibfield  {author} {\bibinfo {author} {\bibfnamefont {S.}~\bibnamefont
  {Iordanskii}}, \bibinfo {author} {\bibfnamefont {Y.~B.}\ \bibnamefont
  {Lyanda-Geller}}, \ and\ \bibinfo {author} {\bibfnamefont {G.}~\bibnamefont
  {Pikus}},\ }\href@noop {} {\bibfield  {journal} {\bibinfo  {journal} {ZhETF
  Pisma Redaktsiiu}\ }\textbf {\bibinfo {volume} {60}},\ \bibinfo {pages} {199}
  (\bibinfo {year} {1994})}\BibitemShut {NoStop}%
\bibitem [{\citenamefont {Niimi}\ \emph {et~al.}(2010)\citenamefont {Niimi},
  \citenamefont {Baines}, \citenamefont {Capron}, \citenamefont {Mailly},
  \citenamefont {Lo}, \citenamefont {Wieck}, \citenamefont {Meunier},
  \citenamefont {Saminadayar},\ and\ \citenamefont {B\"auerle}}]{Yasuhiro}%
  \BibitemOpen
  \bibfield  {author} {\bibinfo {author} {\bibfnamefont {Y.}~\bibnamefont
  {Niimi}}, \bibinfo {author} {\bibfnamefont {Y.}~\bibnamefont {Baines}},
  \bibinfo {author} {\bibfnamefont {T.}~\bibnamefont {Capron}}, \bibinfo
  {author} {\bibfnamefont {D.}~\bibnamefont {Mailly}}, \bibinfo {author}
  {\bibfnamefont {F.-Y.}\ \bibnamefont {Lo}}, \bibinfo {author} {\bibfnamefont
  {A.~D.}\ \bibnamefont {Wieck}}, \bibinfo {author} {\bibfnamefont
  {T.}~\bibnamefont {Meunier}}, \bibinfo {author} {\bibfnamefont
  {L.}~\bibnamefont {Saminadayar}}, \ and\ \bibinfo {author} {\bibfnamefont
  {C.}~\bibnamefont {B\"auerle}},\ }\href {\doibase 10.1103/PhysRevB.81.245306}
  {\bibfield  {journal} {\bibinfo  {journal} {Phys. Rev. B}\ }\textbf {\bibinfo
  {volume} {81}},\ \bibinfo {pages} {245306} (\bibinfo {year}
  {2010})}\BibitemShut {NoStop}%
\bibitem [{Sup()}]{Suppl}%
  \BibitemOpen
  \href@noop {} {\enquote {\bibinfo {title} {See supplemental material for
  detailed chacteristics of the wafer heterostructure},}\ }\BibitemShut
  {NoStop}%
\bibitem [{\citenamefont {Suzuki}\ \emph {et~al.}(2011)\citenamefont {Suzuki},
  \citenamefont {Harada}, \citenamefont {Maeda}, \citenamefont {Onomitsu},
  \citenamefont {Yamaguchi},\ and\ \citenamefont {Muraki}}]{GateHysteresis}%
  \BibitemOpen
  \bibfield  {author} {\bibinfo {author} {\bibfnamefont {K.}~\bibnamefont
  {Suzuki}}, \bibinfo {author} {\bibfnamefont {Y.}~\bibnamefont {Harada}},
  \bibinfo {author} {\bibfnamefont {F.}~\bibnamefont {Maeda}}, \bibinfo
  {author} {\bibfnamefont {K.}~\bibnamefont {Onomitsu}}, \bibinfo {author}
  {\bibfnamefont {T.}~\bibnamefont {Yamaguchi}}, \ and\ \bibinfo {author}
  {\bibfnamefont {K.}~\bibnamefont {Muraki}},\ }\href
  {http://stacks.iop.org/1882-0786/4/i=12/a=125702} {\bibfield  {journal}
  {\bibinfo  {journal} {Applied Physics Express}\ }\textbf {\bibinfo {volume}
  {4}},\ \bibinfo {pages} {125702} (\bibinfo {year} {2011})}\BibitemShut
  {NoStop}%
\bibitem [{\citenamefont {Shibata}\ \emph {et~al.}(2019)\citenamefont
  {Shibata}, \citenamefont {Karalic}, \citenamefont {Mittag}, \citenamefont
  {Tschirky}, \citenamefont {Reichl}, \citenamefont {Ito}, \citenamefont
  {Hashimoto}, \citenamefont {Tomimatsu}, \citenamefont {Hirayama},
  \citenamefont {Wegscheider}, \citenamefont {Ihn},\ and\ \citenamefont
  {Ensslin}}]{GateHysteresis2}%
  \BibitemOpen
  \bibfield  {author} {\bibinfo {author} {\bibfnamefont {K.}~\bibnamefont
  {Shibata}}, \bibinfo {author} {\bibfnamefont {M.}~\bibnamefont {Karalic}},
  \bibinfo {author} {\bibfnamefont {C.}~\bibnamefont {Mittag}}, \bibinfo
  {author} {\bibfnamefont {T.}~\bibnamefont {Tschirky}}, \bibinfo {author}
  {\bibfnamefont {C.}~\bibnamefont {Reichl}}, \bibinfo {author} {\bibfnamefont
  {H.}~\bibnamefont {Ito}}, \bibinfo {author} {\bibfnamefont {K.}~\bibnamefont
  {Hashimoto}}, \bibinfo {author} {\bibfnamefont {T.}~\bibnamefont
  {Tomimatsu}}, \bibinfo {author} {\bibfnamefont {Y.}~\bibnamefont {Hirayama}},
  \bibinfo {author} {\bibfnamefont {W.}~\bibnamefont {Wegscheider}}, \bibinfo
  {author} {\bibfnamefont {T.}~\bibnamefont {Ihn}}, \ and\ \bibinfo {author}
  {\bibfnamefont {K.}~\bibnamefont {Ensslin}},\ }\href {\doibase
  10.1063/1.5093133} {\bibfield  {journal} {\bibinfo  {journal} {Applied
  Physics Letters}\ }\textbf {\bibinfo {volume} {114}},\ \bibinfo {pages}
  {232102} (\bibinfo {year} {2019})}\BibitemShut {NoStop}%
\bibitem [{\citenamefont {Altshuler}\ \emph {et~al.}(1982)\citenamefont
  {Altshuler}, \citenamefont {Aronov},\ and\ \citenamefont
  {Khmelnitsky}}]{Altshuler}%
  \BibitemOpen
  \bibfield  {author} {\bibinfo {author} {\bibfnamefont {B.~L.}\ \bibnamefont
  {Altshuler}}, \bibinfo {author} {\bibfnamefont {A.~G.}\ \bibnamefont
  {Aronov}}, \ and\ \bibinfo {author} {\bibfnamefont {D.~E.}\ \bibnamefont
  {Khmelnitsky}},\ }\href {\doibase 10.1088/0022-3719/15/36/018} {\bibfield
  {journal} {\bibinfo  {journal} {Journal of Physics C: Solid State Physics}\
  }\textbf {\bibinfo {volume} {15}},\ \bibinfo {pages} {7367} (\bibinfo {year}
  {1982})}\BibitemShut {NoStop}%
\bibitem [{\citenamefont {Taboryski}\ and\ \citenamefont
  {Lindelof}(1990)}]{Taboryski_1990}%
  \BibitemOpen
  \bibfield  {author} {\bibinfo {author} {\bibfnamefont {R.}~\bibnamefont
  {Taboryski}}\ and\ \bibinfo {author} {\bibfnamefont {P.~E.}\ \bibnamefont
  {Lindelof}},\ }\href {\doibase 10.1088/0268-1242/5/9/003} {\bibfield
  {journal} {\bibinfo  {journal} {Semiconductor Science and Technology}\
  }\textbf {\bibinfo {volume} {5}},\ \bibinfo {pages} {933} (\bibinfo {year}
  {1990})}\BibitemShut {NoStop}%
\end{thebibliography}%


\begin{thebibliography}{3}%
\makeatletter
\providecommand \@ifxundefined [1]{%
 \@ifx{#1\undefined}
}%
\providecommand \@ifnum [1]{%
 \ifnum #1\expandafter \@firstoftwo
 \else \expandafter \@secondoftwo
 \fi
}%
\providecommand \@ifx [1]{%
 \ifx #1\expandafter \@firstoftwo
 \else \expandafter \@secondoftwo
 \fi
}%
\providecommand \natexlab [1]{#1}%
\providecommand \enquote  [1]{``#1''}%
\providecommand \bibnamefont  [1]{#1}%
\providecommand \bibfnamefont [1]{#1}%
\providecommand \citenamefont [1]{#1}%
\providecommand \href@noop [0]{\@secondoftwo}%
\providecommand \href [0]{\begingroup \@sanitize@url \@href}%
\providecommand \@href[1]{\@@startlink{#1}\@@href}%
\providecommand \@@href[1]{\endgroup#1\@@endlink}%
\providecommand \@sanitize@url [0]{\catcode `\\12\catcode `\$12\catcode
  `\&12\catcode `\#12\catcode `\^12\catcode `\_12\catcode `\%12\relax}%
\providecommand \@@startlink[1]{}%
\providecommand \@@endlink[0]{}%
\providecommand \url  [0]{\begingroup\@sanitize@url \@url }%
\providecommand \@url [1]{\endgroup\@href {#1}{\urlprefix }}%
\providecommand \urlprefix  [0]{URL }%
\providecommand \Eprint [0]{\href }%
\providecommand \doibase [0]{http://dx.doi.org/}%
\providecommand \selectlanguage [0]{\@gobble}%
\providecommand \bibinfo  [0]{\@secondoftwo}%
\providecommand \bibfield  [0]{\@secondoftwo}%
\providecommand \translation [1]{[#1]}%
\providecommand \BibitemOpen [0]{}%
\providecommand \bibitemStop [0]{}%
\providecommand \bibitemNoStop [0]{.\EOS\space}%
\providecommand \EOS [0]{\spacefactor3000\relax}%
\providecommand \BibitemShut  [1]{\csname bibitem#1\endcsname}%
\let\auto@bib@innerbib\@empty
\bibitem [{\citenamefont {Sazgari}\ \emph {et~al.}(2019)\citenamefont
  {Sazgari}, \citenamefont {Sullivan},\ and\ \citenamefont {Kaya}}]{Sazgari}%
  \BibitemOpen
  \bibfield  {author} {\bibinfo {author} {\bibfnamefont {V.}~\bibnamefont
  {Sazgari}}, \bibinfo {author} {\bibfnamefont {G.}~\bibnamefont {Sullivan}}, \
  and\ \bibinfo {author} {\bibfnamefont {I.~I.}\ \bibnamefont {Kaya}},\ }\href
  {\doibase 10.1103/PhysRevB.100.041404} {\bibfield  {journal} {\bibinfo
  {journal} {Phys. Rev. B}\ }\textbf {\bibinfo {volume} {100}},\ \bibinfo
  {pages} {041404} (\bibinfo {year} {2019})}\BibitemShut {NoStop}%
\bibitem [{\citenamefont {Suzuki}\ \emph {et~al.}(2011)\citenamefont {Suzuki},
  \citenamefont {Harada}, \citenamefont {Maeda}, \citenamefont {Onomitsu},
  \citenamefont {Yamaguchi},\ and\ \citenamefont {Muraki}}]{GateHysteresis}%
  \BibitemOpen
  \bibfield  {author} {\bibinfo {author} {\bibfnamefont {K.}~\bibnamefont
  {Suzuki}}, \bibinfo {author} {\bibfnamefont {Y.}~\bibnamefont {Harada}},
  \bibinfo {author} {\bibfnamefont {F.}~\bibnamefont {Maeda}}, \bibinfo
  {author} {\bibfnamefont {K.}~\bibnamefont {Onomitsu}}, \bibinfo {author}
  {\bibfnamefont {T.}~\bibnamefont {Yamaguchi}}, \ and\ \bibinfo {author}
  {\bibfnamefont {K.}~\bibnamefont {Muraki}},\ }\href
  {http://stacks.iop.org/1882-0786/4/i=12/a=125702} {\bibfield  {journal}
  {\bibinfo  {journal} {Applied Physics Express}\ }\textbf {\bibinfo {volume}
  {4}},\ \bibinfo {pages} {125702} (\bibinfo {year} {2011})}\BibitemShut
  {NoStop}%
\bibitem [{\citenamefont {Shibata}\ \emph {et~al.}(2019)\citenamefont
  {Shibata}, \citenamefont {Karalic}, \citenamefont {Mittag}, \citenamefont
  {Tschirky}, \citenamefont {Reichl}, \citenamefont {Ito}, \citenamefont
  {Hashimoto}, \citenamefont {Tomimatsu}, \citenamefont {Hirayama},
  \citenamefont {Wegscheider}, \citenamefont {Ihn},\ and\ \citenamefont
  {Ensslin}}]{GateHysteresis2}%
  \BibitemOpen
  \bibfield  {author} {\bibinfo {author} {\bibfnamefont {K.}~\bibnamefont
  {Shibata}}, \bibinfo {author} {\bibfnamefont {M.}~\bibnamefont {Karalic}},
  \bibinfo {author} {\bibfnamefont {C.}~\bibnamefont {Mittag}}, \bibinfo
  {author} {\bibfnamefont {T.}~\bibnamefont {Tschirky}}, \bibinfo {author}
  {\bibfnamefont {C.}~\bibnamefont {Reichl}}, \bibinfo {author} {\bibfnamefont
  {H.}~\bibnamefont {Ito}}, \bibinfo {author} {\bibfnamefont {K.}~\bibnamefont
  {Hashimoto}}, \bibinfo {author} {\bibfnamefont {T.}~\bibnamefont
  {Tomimatsu}}, \bibinfo {author} {\bibfnamefont {Y.}~\bibnamefont {Hirayama}},
  \bibinfo {author} {\bibfnamefont {W.}~\bibnamefont {Wegscheider}}, \bibinfo
  {author} {\bibfnamefont {T.}~\bibnamefont {Ihn}}, \ and\ \bibinfo {author}
  {\bibfnamefont {K.}~\bibnamefont {Ensslin}},\ }\href {\doibase
  10.1063/1.5093133} {\bibfield  {journal} {\bibinfo  {journal} {Applied
  Physics Letters}\ }\textbf {\bibinfo {volume} {114}},\ \bibinfo {pages}
  {232102} (\bibinfo {year} {2019})}\BibitemShut {NoStop}%
\end{thebibliography}%
\end{document}


\title{Supplemental Material for 'Interaction-induced crossover between weak anti-localization and weak localization in a disordered InAs/GaSb double quantum well'}

\author{Vahid Sazgari}
\affiliation{Faculty of Engineering and Natural Sciences, Sabanci University, Tuzla, 34956 Istanbul, Turkey}
\affiliation{Sabanci University Nanotechnology Research and Application Center, Tuzla, 34956 Istanbul, Turkey}
\author{Gerard Sullivan}
\affiliation{Teledyne Scientific and Imaging, Thousand Oaks, CA 91630, USA}
\author{\.{I}smet \.{I} Kaya}
\affiliation{Faculty of Engineering and Natural Sciences, Sabanci University, Tuzla, 34956 Istanbul, Turkey}
\affiliation{Sabanci University Nanotechnology Research and Application Center, Tuzla, 34956 Istanbul, Turkey}

 \email{iikaya@sabanciuniv.edu}

\date{\today}
\maketitle

\begin{figure}[ht]
	\centering
	\includegraphics[width=0.6\textwidth]{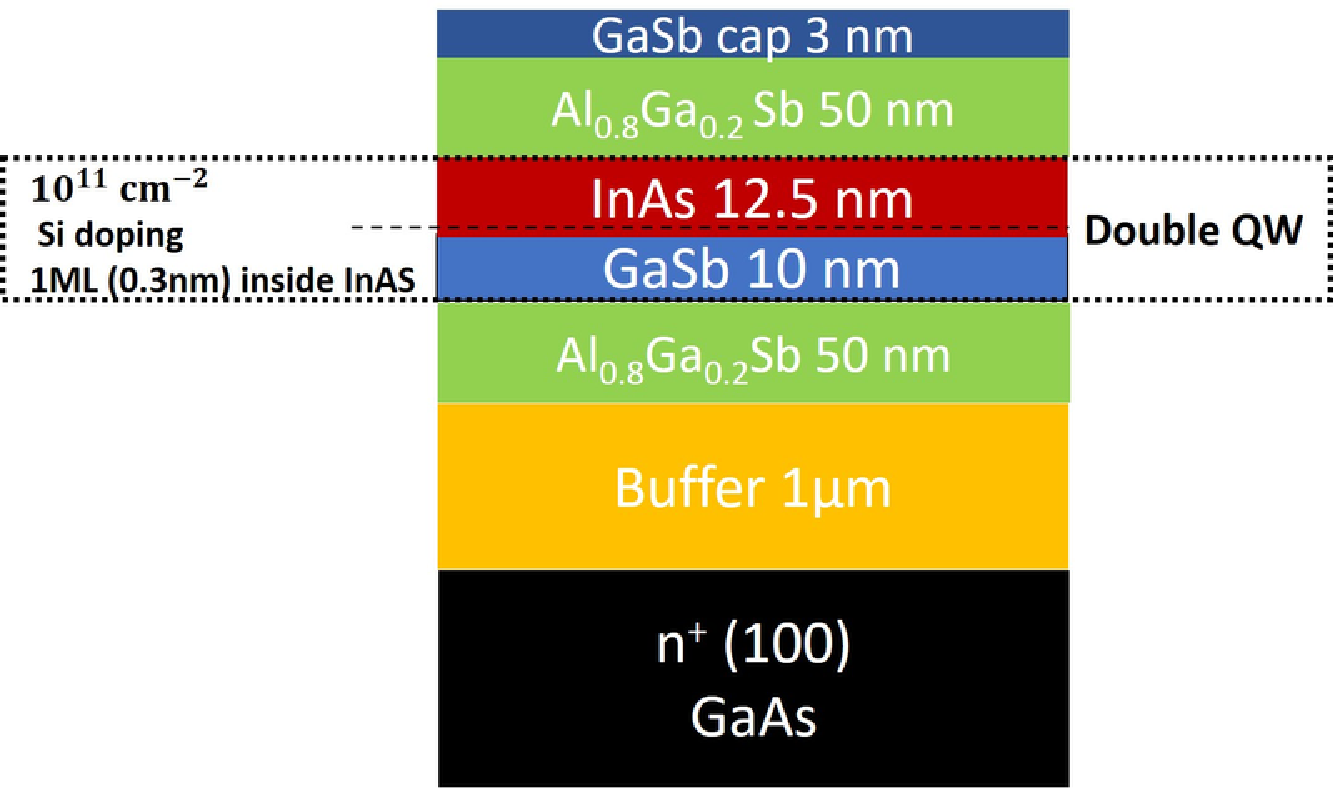}
	\caption {Schematic of the wafer heterostructure studied in this work with Si delta-dopants inside InAs QW at 1 monolayer distance from the InAs/GaSb interface..}\label{FIG:Wafer}
\end{figure}

The disordered bilayer InAs/GaSb bilayer quantum well structure is described in Fig.~\ref{FIG:Wafer}. A semi-metallic 2D bulk transport is confirmed by resistance measurement of a Hall bar device and a Corbino geometry. Both devices are characterized by resistance plateaus of different values at lowered top gate voltages. Since the Corbino device has a relatively small resistance, it is hard to estimate the contact resistance. Therefore, we compare the change of resistivity as the top gate voltage is swept from positive to negative values depleting the electrons.

\begin{figure}
	\centering
	\includegraphics[width=\textwidth]{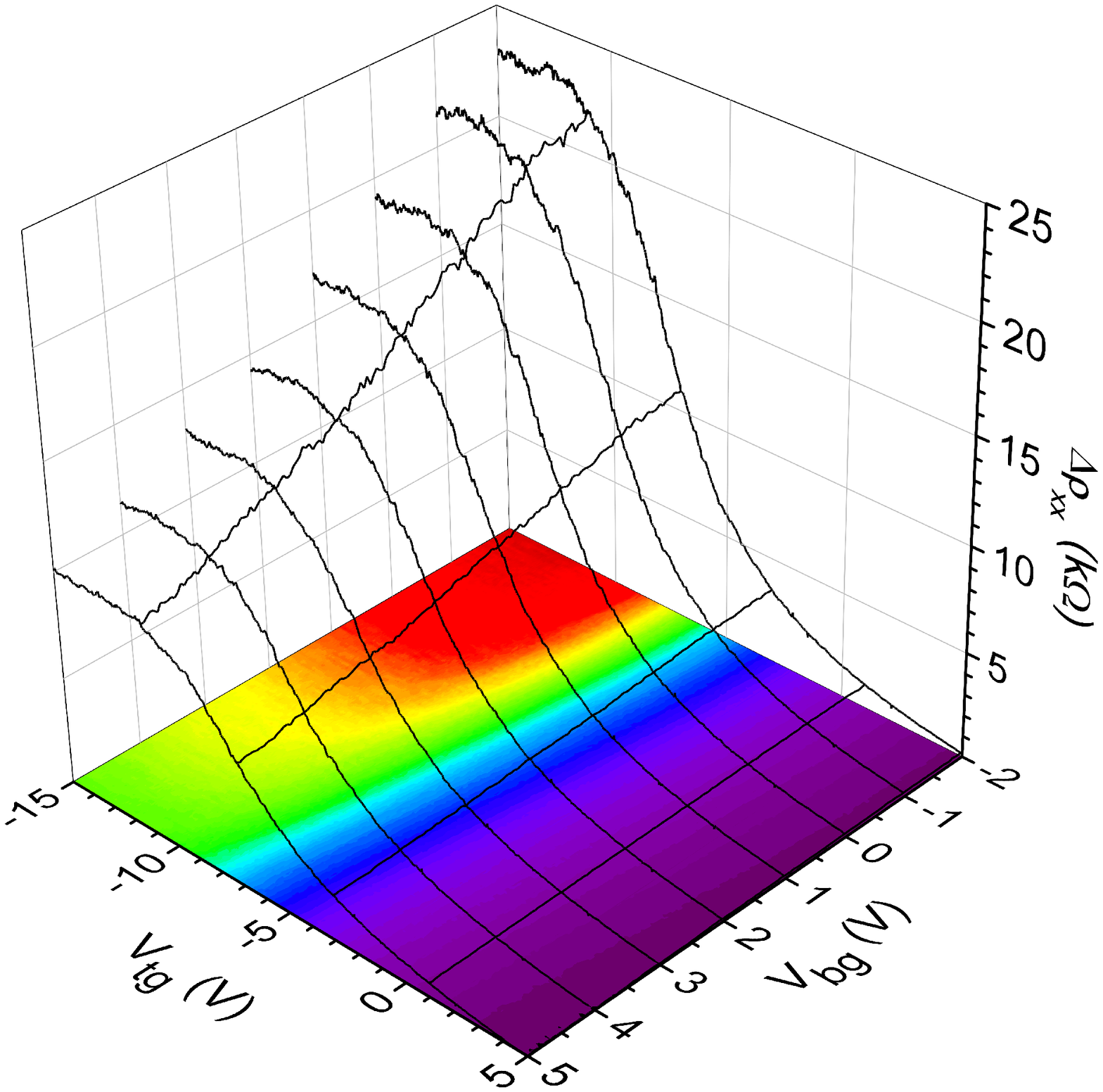}
	\caption {The relative change of the longitudinal resistivity of the Hall bar device as a function of $V_{tg}$ and $V_{bg}$ measured at $T~=~10$~mK with 10~nA DC current. The sample exhibits semi-metallic behavior in the whole range of top and bottom gate voltages.}\label{FIG:W1864-Phase}
\end{figure}

The change of the longitudinal resistivity of the Hall bar device with respect to the top and bottom gate voltages is shown in Fig.~\ref{FIG:W1864-Phase}. It is noticeable that for all bottom gate voltages, the resistivity saturates as $V_{tg}$ is lowered and forms plateaus of different values when $V_{tg}\le -12~$~V. 

\begin{figure}
	\centering
	\includegraphics[width=0.8\textwidth]{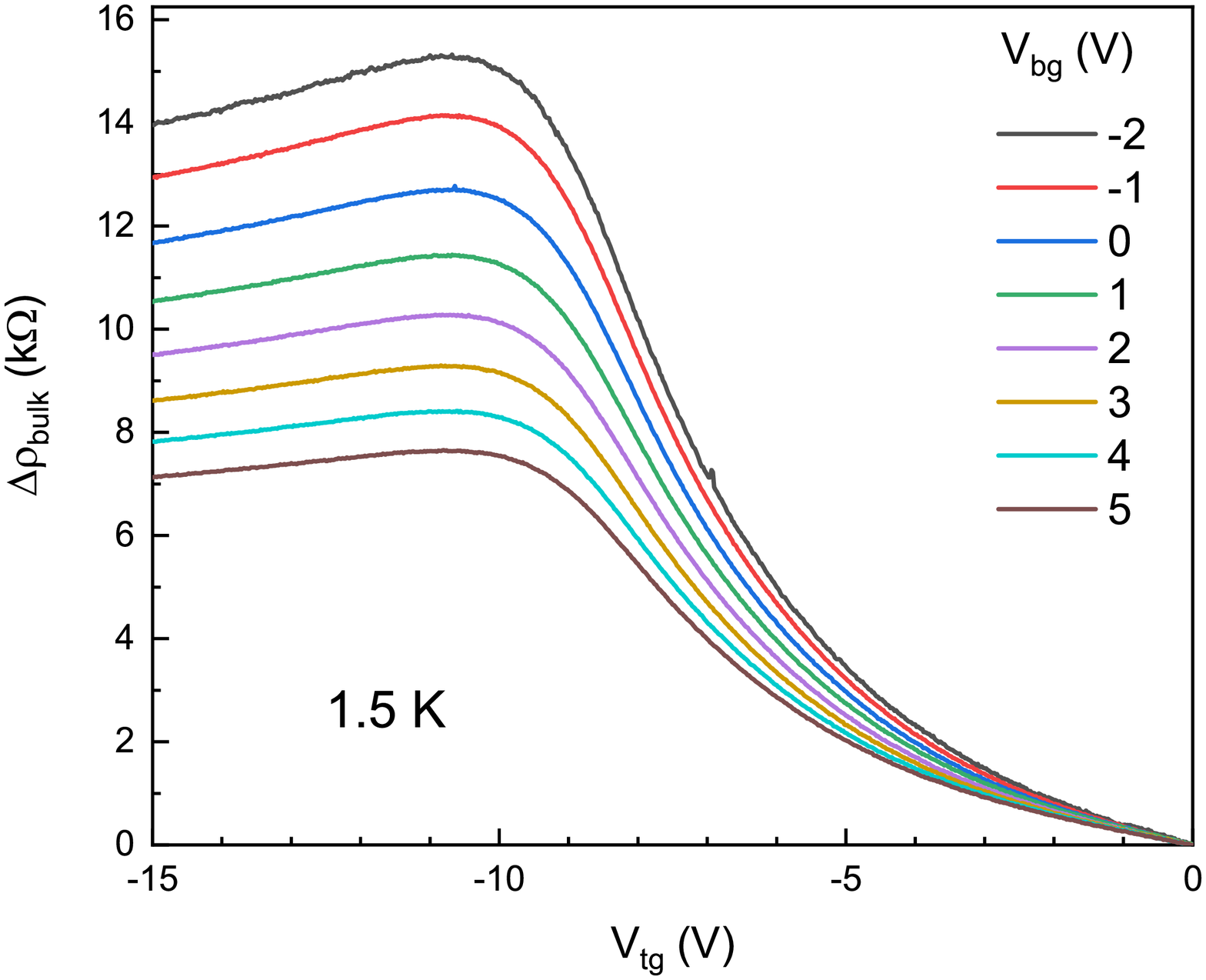}
	\caption {The change of resistivity for the Corbino device as a function of $V_{tg}$ for different $V_{bg}$ measured with 100~$\mu V$ AC current between source and drain at $T=~1.5$~K. The relative modulation of resistivity is comparable with that of the Hall bar.}\label{FIG:W1864-Corbino}
\end{figure}

To study the bulk transport behavior, a Corbino disk which is top-gated by a ring-shaped metal with inner and outer diameters of $r_i$~=~400~$\mu$m and $r_o$~=~600~$\mu$m respectively were used in this study. The same behavior is observed in the Corbino device which indicates that the plateaus are not related to the edge transport behavior. The bulk resistivity of the Corbino device (Fig.~\ref{FIG:W1864-Corbino}) is consistent with the Hall bar resistivity, which confirms that the transport in the Hall bar device is mediated by the bulk carriers.  

The density modulation with respect to the top gate electric field and formation of plateaus imply that the QW channel can not be fully depleted. We do not fully understand this behavior, however, it may be attributed to the surface trap states common to the III-V semiconductor interfaces~\cite{Sazgari,GateHysteresis,GateHysteresis2} or impurity-induced charge states inside the bilayer QW. 

\bibliographystyle{apsrev4-1}
\bibliography{Supplemental-WAL_WL}